\documentclass{aastex63}
\usepackage{CJK}

\newcommand\rmi{\mathrm{i}}
\newcommand{\comment}[1]{}

\shorttitle{Drag Instability}
\begin{document}
%\begin{CJK*}{UTF8}{gbsn}
%\begin{CJK*}{UTF8}{bsmi}
%\title{Template \aastex Article with Examples: 
%v6.3\footnote{Released on June, 10th, 2019}}
\title{The Drag Instability in a 1D Isothermal C-Shock}

\correspondingauthor{Pin-Gao Gu}
\email{gu@asiaa.sinica.edu.tw}

%\author{Pin-Gao Gu (辜品高)}
\author{Pin-Gao Gu}
\affiliation{Institute of Astronomy \& Astrophysics, Academia Sinica,
%1667 K Street NW, Suite 800 \\
Taipei 10617, Taiwan}

%\author{Che-Yu Chen (陳哲聿)}
\author{Che-Yu Chen}
\affiliation{Department of Astronomy, University of Virginia, Charlottesville, VA 22904, USA}
%1667 K Street NW, Suite 800 \\
%Taipei 10617, Taiwan}
%\affiliation{Department of Physics, National Taiwan University\\
%Taipei 10617, Taiwan}
%1667 K Street NW, Suite 800 \\
%Washington, DC 20006, USA}

%\author{Chien-Chang Yen}
%\affiliation{Institute of Astronomy \& Astrophysics, Academia Sinica\\
%1667 K Street NW, Suite 800 \\
%Taipei 10617, Taiwan}
%\affiliation{Department of Mathematics, Fu-Jen Catholic University\\
%New Taipei City 24205, Taiwan}

%\collaboration{1}{(AAS Journals Data Scientists collaboration)}

%\author{Butler Burton}
%\affiliation{Leiden University}
%\affiliation{AAS Journals Associate Editor-in-Chief}
%\nocollaboration{1}

%\nocollaboration{2}

%% Note that the \and command from previous versions of AASTeX is now
%% depreciated in this version as it is no longer necessary. AASTeX 
%% automatically takes care of all commas and "and"s between authors names.

%% AASTeX 6.3 has the new \collaboration and \nocollaboration commands to
%% provide the collaboration status of a group of authors. These commands 
%% can be used either before or after the list of corresponding authors. The
%% argument for \collaboration is the collaboration identifier. Authors are
%% encouraged to surround collaboration identifiers with ()s. The 
%% \nocollaboration command takes no argument and exists to indicate that
%% the nearby authors are not part of surrounding collaborations.

%% Mark off the abstract in the ``abstract'' environment. 
\begin{abstract}
We investigate whether the drag instability, proposed by Gu et al., occurs in a one-dimensional (1D) C-shock. The 1D background model proposed by Chen \& Ostriker for a steady isothermal C-shock is adopted, and a 1D isothermal linear analysis is performed. We confirm the postulation of Gu et al. that the drift velocity between the ions and the neutrals is sufficiently high within a C-shock to allow for the drag instability. We also study the underlying physics of the decaying modes in the shock and post-shock regions.  The drag instability is an overstability phenomenon associated with an exponentially growing mode of a propagating wave. We find that the growing wave mode can only propagate downstream within the shock and subsequently decay in the post-shock region. The maximum total growth (MTG) for such an unstable wave before it is damped is estimated in typical environments of star-forming clouds, which is
approximately 10-30 times larger than the initial perturbation at the modest shock velocities and can be significantly enhanced several hundred times for a stronger C-shock with a larger width.
\end{abstract}

%% Keywords should appear after the \end{abstract} command. 
%% See the online documentation for the full list of available subject
%% keywords and the rules for their use.
%\keywords{editorials, notices --- 
%miscellaneous --- catalogs --- surveys}

%% From the front matter, we move on to the body of the paper.
%% Sections are demarcated by \section and \subsection, respectively.
%% Observe the use of the LaTeX \label
%% command after the \subsection to give a symbolic KEY to the
%% subsection for cross-referencing in a \ref command.
%% You can use LaTeX's \ref and \label commands to keep track of
%% cross-references to sections, equations, tables, and figures.
%% That way, if you change the order of any elements, LaTeX will
%% automatically renumber them.
%%
%% We recommend that authors also use the natbib \citep
%% and \citet commands to identify citations.  The citations are
%% tied to the reference list via symbolic KEYs. The KEY corresponds
%% to the KEY in the \bibitem in the reference list below. 

%\section{Introduction} \label{sec:intro}

\section{Introduction}
\label{sec:intro}
Stars form within the molecular clouds \citep{Shu1987}, which are the densest subregions of the interstellar medium (ISM). 
While the galactic star formation efficiency is heavily regulated by thermal and dynamic feedback from young stars \citep[see, e.g.,][]{Ostriker10},
it is widely recognized that 
%the star formation process can be significantly modified by the interstellar magnetic field 
the interstellar magnetic field plays a critical role in modifying the star formation process locally within individual clouds
\citep{MO07,Crutcher12}. 
However, 
%In these cold molecular clouds and their substructures (temperature $\sim 10$~K; see e.g.,~\citealt{Fukui10}), the gas is generally weakly ionized because of the relatively low ionization rate by cosmic rays ($\xi_\mathrm{CR} \sim 10^{-17}~\mathrm{s}^{-1}$; see e.g.,~\citealt{Draine, Indriolo12}). The star-forming gas is therefore mainly composed of the neutrals with only a small abundance of ions, with typical ionization fraction $\lesssim 10^{-6}$
the gas in these cold molecular clouds and their substructures is generally weakly ionized
\citep[see e.g.,][]{Tielens05, Dalgarno06},
%The ions may be considered to be tightly coupled with interstellar magnetic fields. 
%Consequently, 
and
%the collisional coupling between neutrals and ions plays a critical role in controlling the ability of magnetic fields to affect star formation 
the actual ability of magnetic fields to affect star formation thus relies on the collisional coupling between neutrals and ions
\citep{Mouschovias79}.

With the existence of a spatial gradient of the field lines to exert a Lorentz force on the ions, the ions can drift relative to the neutrals. As the result, the ambipolar diffusion occurs when the drag force (proportional to the ion-neutral collision rate; \citealt{Spitzer1956}) is balanced by the Lorentz force, leading to the diffusion of the magnetic fields from the neutrals \citep{Shu}. 
This allows the redistribution of neutral gas relative to the magnetic flux \citep{MS1956}.
Ambipolar diffusion has been considered as the main mechanism for several processes during star formation, including
%overdense subregions within the magnetically-supported molecular clouds to lose magnetic support and collapse gravitationally to form protostellar systems
the collapse of magnetically supported overdense subregions within the molecular clouds
\citep{Mouschovias1978,Nakano1978,LizanoShu1989,Fiedler92, Fiedler93,Oishi06, LiPS08},
%There have hence been numerous studies on the sequential gravitational condensation induced by ambipolar diffusion within the molecular clouds \citep[e.g.,][]{Nakano1978, Oishi06, Fiedler92, Fiedler93,LiPS08}.
the enhanced angular momentum transport (compared to that in the magnetic braking catastrophe) during protostellar disk formation \citep{MellonLi09,Dapp2012,Hennebelle16,Masson16,Vaytet18,Lam19}, and the development of substructures in protoplanetary disks \citep{BaiStone11,Lesur14,Gressel15,RiolsLesur18,Suriano18,Suriano19}.

Alternatively, 
Gu et al. 2004 (hereafter \citet{Gu04}) studied the stability of ambipolar drift in a weakly ionized fluid. GLV simplified the problem by representing it as that of a 1D drift flow threaded with perpendicular magnetic fields. \citet{Gu04} discovered a local overstable mode provided that the ion-neutral drift velocity $v_{\rm d} \equiv |v_{\rm ion} - v_{\rm neutral}|$ is as high as the Alfv\'en velocity of the bulk fluid ($V_{\rm A} \equiv B/\sqrt{4\pi\rho}$), and if the ionization equilibrium can be sustained. Although a high drift velocity arises from a strong Lorentz force, the instability in its simplest form is not related to any magnetosonic or acoustic modes, but is caused solely by the pronounced drag force induced by the high drift velocity. Consequently, \citet{Gu04} named the instability ``drag instability." We use this same terminology to refer to the overstability in this paper. 

In general, the drag caused by the drift motion between two fluids alone (i.e.,~independent of Alfv\'en, magnetosonic, or acoustic modes) can provide a free energy to facilitate a fluid instability under favorable conditions. A notable example of this phenomenon is the streaming instability caused by the dust-gas drift motion in a protoplanetary disk \citep[e.g.,][]{YG05}. Because the drag instability requires a high ambipolar drift velocity, \citet{Gu04} postulated that the instability occurs in regions where magnetic fields are highly stressed, including interstellar shock systems and/or collapsing protostellar cores. 
While there is evidence that the ion-neutral drift velocity within collapsing protostellar envelopes could be as high as the freefall velocity ($\sim 1$~km~s$^{-1}$; see, e.g.,~\citealt{Yen18,Lam19}),
in this study we focus on interstellar shock systems with efficient ambipolar diffusion to investigate whether the drag instability can take place during the compression that initiates star formation.

In the ISM, stressed magnetic fields may occur due to shock compression triggered by clump-clump collision or the supersonic, turbulent gas flows within the giant molecular clouds \citep[e.g.,][]{MS1956, DM93, Ostriker99, BP07, Federrath11, LiHB14}. 
%such as  a C shock front, a collapsing proto-stellar cloud, or colliding clouds. 
While jump-type (J-type) shocks exhibit sharp supersonic (or super-Alfv\'enic for magnetized shocks) discontinuities in physical properties,
in the case of nonideal magnetohydrodynamics (MHD), continuous-type (C-type) shocks manifest as a smooth transition between the pre- and post-shock regions due to the ambipolar drift between ions and neutrals. Specifically, when the ion-neutral drift velocity is lower than the Alfv\'en speed of the ions, the Alfv\'en speed of the ions can propagate the shock signal upstream, thereby compressing the ions and magnetic fields and subsequently dragging and compressing the neutrals. This process smoothens out the sharp transition and results in a width of the continuous shock profile between the pre- and post-shock regions \citep{Draine80,DM93}.

In the cold molecular clouds and their substructures (temperature $\sim 10$~K; see e.g.,~\citealt{Fukui10}), the ionization rate by cosmic rays is relatively low ($\xi_\mathrm{CR} \sim 10^{-17}~\mathrm{s}^{-1}$; see, e.g.,~\citealt{Draine, Indriolo12}). The star-forming gas is therefore mainly composed of the neutrals with only a small abundance of ions, with typical ionization fraction $\lesssim 10^{-6}$ \citep{Tielens05, Dalgarno06}.
Such weakly ionized gas, when compressed by supersonic turbulence, provides the favored conditions for large ion-neutral drifts and C-type shocks. There have been various observational efforts to probe such features in turbulent molecular clouds \citep[e.g.,][]{LiHoude08,Hezareh10,Hezareh14,XuLi16,Tang18}, although most of these observations are indirect measurements and highly dependent on the adopted dynamical and chemical models \citep[e.g.,][]{Flower98,Flower10,Gusdorf08,LehmannWardle16,Valdivia17}. Theoretically,
previous studies have investigated in detail the formation as well as the physical and chemical properties of C-shocks \citep[e.g.,][]{Wardle, MacLow95, SmithML97,Pineau97,Guillet11}.
In particular, Chen \& Ostriker 2012 (hereafter \citet{CO12}) studied the 1D isothermal C-shock along the drift direction with a transverse magnetic field. They analytically derived the 1D structure of a C-shock, thereby providing an appropriate and convenient background state for the 1D linear analysis of the drag instability proposed in \citet{Gu04}.

We note that while we focus on C-shock instability in this study, the drag instability could exist in other systems with enhanced drift velocities (see \citet{Gu04}).
Moreover, the drag instability differs from the Wardle instability, originally proposed for C-shock systems in \cite{Wardle}, which is analogous to the Park instability, with the ion-neutral drag playing a role of gravity to collect matters in the magnetic ``valley." The Wardle instability therefore requires the wiggle of 2D or 3D field lines and exists as a more global mode along the shock direction. In contrast, the drag instability can exist in a 1D flow and is a local effect.

%and thus differs from the Wardle instability in C shocks \citep{Wardle}.} \textbf{\textcolor{blue}{Furthermore, the Wardle instability is analogues to the Park instability with the ion-neutral drag playing a role of gravity to collect matters in the magnetic ``valley". Therefore, it requires the wiggle of 2D/3D field lines and exists in a form of a more global mode along the shock direction. In contrast, the drag instability can exist in a 1D flow and is a local instability.}}
Furthermore, among all previous investigations of the 1D C-shock structure \citep[e.g.,][]{SmithML97,Chieze98,Ciolek02,vanLoo09,Ashmore10}, only the simplified scenario discussed in \citet{CO12} (isothermal gas with ionization-recombination equilibrium) provides a suitable condition for the drag instability to occur. 
We further note that 
the drag instability differs from the fragmentation instability \citep{Zweibel98}, 
%which occurs in a self-gravitating cloud in both magneto-hydrostatic and ionization equilibrium. 
which requires the system to be near marginal
dynamical stability so that the release of energy through diffusion of the magnetic field could lead to runaway contraction of an initially overdense region. 
The drag instability, on the other hand, does not require hydrostatic equilibrium, and the ultimate source of energy comes from the stressed magnetic fields. 

The outline of the paper is as follows. In \S\ref{sec:linear}, we first review the steady-state C-shock solution of \citet{CO12} and the linear theory of \citet{Gu04}. 
%By considering a fiducial C-shock model as the background state, the dispersion relation for the drag instability is summarized as well as the other dispersion relations in the post-shock region is presented.
By considering a fiducial C-shock model as the background state (\S\ref{sec:model}), we present the dispersion relation for the drag instability and the other dispersion relations in the postshock region (\S\ref{sec:dispersion}). 
%Other dispersion relations in the post-shock region are also presented. 
The exact solutions of the eigenvalues and eigenmodes are obtained by solving the linearized equations and are analyzed by using the dispersion relations (\S\ref{sec:eigen}). In \S\ref{sec:growth}, we present the maximum total growth (MTG) obtained for the unstable mode under the drag instability within a C-shock by using the fiducial model and other models with different pre-shock conditions. 
We discuss the connection between this analytic work and previous numerical time-dependent simulations in \S\ref{sec:disc}.
Finally, the results of this study are summarized in \S\ref{sec:sum}.

\section{Linear analysis: WKBJ analysis}
\label{sec:linear}

In general, the dynamical evolution of ions and neutrals is governed by their individual continuity and momentum equations,
in addition to the collisional drag force, cosmic-ray ionization, ion-electron recombination in the gas phase, and 
the induction equation for ions \citep[e.g.,][]{Draine80,Shu,CO12}. The equations are as follows.
\begin{eqnarray}
{\partial \rho_n \over \partial t}+ \nabla \cdot (\rho_n {\bf v_n})=0, \label{eq:1}\\
{\partial \rho_i \over \partial t}+ \nabla \cdot (\rho_i {\bf v_i})= - \beta \rho_i^2 + \xi_\mathrm{CR} \rho_n, \label{eq:2} \\
\rho_n \left[ {\partial {\bf v_n} \over \partial t} + ({\bf v_n} \cdot \nabla ) {\bf v_n} \right]+\nabla p_n ={\bf f_d}, \label{eq:3}\\
\rho_i \left[ {\partial {\bf v_i} \over \partial t} + ({\bf v_i} \cdot \nabla ) {\bf v_i} \right]+\nabla p_i -{1\over 4\pi}(\nabla \times {\bf B})\times {\bf B}=-{\bf f_d}, \\
{\partial {\bf B} \over \partial t} + \nabla \times ({\bf B} \times {\bf v_i})=0, \label{eq:5}
\end{eqnarray}
where ${\bf v}$ is the velocity, $\rho$ is the density, ${\bf B}$ is the magnetic field, $p=\rho c_s^2$ is the gas pressure when the isothermal sound speed $c_s$ is 0.2~km/s at a temperature $\sim 10$~K, and the subscripts $i$ and $n$ denote the ion and neutral species, respectively. 
Note that the neutrals and ions are coupled by the collisional drag force ${\bf f_d}\equiv \gamma \rho_i \rho_n {\bf v_d}=\gamma \rho_i \rho_n ({\bf v_i}-{\bf v_n})$, where $\gamma \approx 3.5 \times 10^{13}$~cm$^3$~s$^{-1}$~g$^{-1}$ is the drag force coefficient \citep{Draine}. 
The evolution of ion number density is controlled by the cosmic-ray ionization rate $\xi_\mathrm{CR}$ and the ion recombination in the gas phase $\beta$ \citep[see, e.g.,][]{CO12}.

%Here, $\beta \approx 10^{-7}$~cm$^3$~s$^{-1}/m_i$ is the recombination rate coefficient, the ionization rate due to cosmic rays is $\xi \approx 10^{-17}$~s$^{-1}$~($m_i/m_n$), $\gamma \approx 3.5 \times 10^{13}$~cm$^3$~s$^{-1}$~g$^{-1}$ is the drag force coefficient \citep{Draine}, and ${\bf f_d}=\gamma \rho_i \rho_n ({\bf v_i}-{\bf v_n}) \equiv \gamma \rho_i \rho_n {\bf v_d}$ is the drag force with the drift velocity $\bf v_d$. 
%Given the ionization and recombination rates, the parameter $\chi_{i0}$ in the expression $n_i=10^{-6} \chi_{i0}n_n^{1/2}$ of \citet{CO12}  in the ionization-recombination equilibrium is then given by $10^6 \sqrt{\xi (m_n/m_i)/(\beta m_i)} =10$, which falls in the typical range of $\chi_{i0}$ \citep{McKee10}. 

\subsection{Background States and Linearized Equations}

%We adopt the 1D steady-state C-shock model shown in Figure 3 of \citet{CO12} as the background state in the shock frame, with the pre-shock parameters 
%$n_0=500$ cm$^{-3}$, $v_0=5$ km/s, $B_0=5 \mu$G, and $\chi_{i0}=10$ (see their Figure 2).  
%$n_0=500$ cm$^{-3}$ (neutral number density), $V_{i,0}=V_{n,0}=v_0=5$ km/s (shock velocity), $B_0=10 \mu$G and $\chi_{i0}=10$ to be our fiducial model. The choice of the fiducial model is made to be reproduced easily by comparing to the existing results in \citet{CO12}.
Following the simplified scenario discussed in \cite{CO12} that the magnetic field is perpendicular to the gas flow toward the $+x$ direction through the shock, 
the equilibrium equations in this 1D C-shock system are given by
\begin{eqnarray}
V_n {d \rho_n \over dx} = - \rho_n {d V_n \over dx},\\
V_n  {d V_n \over dx} = \gamma \rho_i V_d - {c_s^2\over \rho_n} {d \rho_n \over dx},\\
\beta \rho_i^2 = \xi_\mathrm{CR} \rho_n,\\
\gamma \rho_n V_d = - V_{A,i}^2 {d \ln B \over dx},\\
V_i {dB \over dx} = - B {d V_i \over dx}.
\end{eqnarray}
In the above equilibrium equations, we consider the strong-coupling approximation under which the ion-neutral drag is balanced by the magnetic pressure gradient for the ions; i.e., $\gamma \rho_n L_B V_d=V_{A,i}^2$ where $L_B \equiv (-d \ln B/dx)^{-1}$. In addition, the equilibrium between cosmic ionization and recombination is assumed; namely, $\beta \rho_i^2 = \xi_\mathrm{CR} \rho_n$ (see \cite{CO12} for justifications of this choice). We note that these equilibrium states are consistent with the background states considered by \citet{Gu04}. 

By subjecting the equilibrium equations to the zero-gradient boundary conditions ($d/dx=0$) far upstream and downstream (i.e., no structures in the steady pre- and post-shock regions), \citet{CO12} derived the 1D structure equation of a C-shock as follows (here and throughout this paper, we use the subscript 0 to denote a physical quantity in the pre-shock region):
\begin{equation}
%{d r_n \over dx}=-{D r_n^{3/2} \over 1 - M^2/r_n^2} \left( {1\over r_n} - {1\over r_B} \right),
{d r_B \over dx}=-{\gamma \rho_{i,0} \over  v_0}\mathcal{M_A}^2{r_n^{3/2} \over r_B} \left( {1\over r_B} -{1\over r_n} \right), \label{eq:drBdx}
\end{equation} 
where the field compression ratio is $r_B \equiv B/B_0 = V_{i,0}/V_i$,
the neutral compression ratio is $r_n \equiv \rho_n/\rho_{n,0}=V_{n,0}/V_n$,
%\left[ 1+2\mathcal{M_A}^2+\beta_{plasma}-r_B^2-\sqrt{[1+2\mathcal{M_A}^2+\beta_{plasma}-r_B^2 ]^2-8\beta_{plasma}\mathcal{M_A}^2} \right]/2\beta_{plasma}$, 
%the field and ion compression ratio $r_B=B/B_0=\rho_i/\rho_{i,0}=V_{i,0}/V_i=[ 1+ \beta_{plasma} (r_n-1)({\mathcal M}^2/r_n-1)]^{1/2}$, $D \equiv (\beta \rho_{i,0}/v_0){\mathcal M}^2$, 
and the Alfv\'en Mach number ${\mathcal M_A}$ for the shock velocity $v_0$ is defined as $ v_0/V_{A,n,0}$. 
Note that $r_n$ can be written as a function of $r_B$:
\begin{equation}
   r_n = \left[ 1+2\mathcal{M_A}^2+\beta_{\rm plasma}-r_B^2-\sqrt{[1+2\mathcal{M_A}^2+\beta_{\rm plasma}-r_B^2 ]^2-8\beta_{\rm plasma}\mathcal{M_A}^2} \right]/2\beta_{\rm plasma},
\end{equation}
with the plasma beta value $\beta_{\rm plasma}\equiv 8\pi \rho_0 c_s^2/B_0^2$.
By placing the shock front at $x=0$, the physical quantities along the C-shock can be obtained by integrating the ordinary different equation (Equation~(\ref{eq:drBdx})) backward from far downstream, where
\begin{equation}
  r_B=r_n=4\mathcal{M_A}^2/[1+\beta_{\rm plasma}+[(1+\beta_{\rm plamsa})^2+8\mathcal{M_A}^2]^{1/2}]
\end{equation}
to far upstream.
In this setup of the problem, the background drift velocity $V_d=V_i-V_n<0$ inside the C-shock (i.e. within the smooth shock transition).

We now consider the perturbations $U(\omega, k)\equiv (\delta \rho_i, \delta v_i, \delta B, \delta \rho_n, \delta v_n)^T$ multiplied by $\exp[\rmi (kx+\omega t)]$ under the Wentzel-Kramers-Brillouin-Jeffreys (WKBJ) approximation. By substituting these perturbations and the background states into the equations (\ref{eq:1})-(\ref{eq:5}), the following linearized equations are obtained \citep{Gu04}:

\comment{
\begin{eqnarray}
-2 \beta \rho_i {\delta \rho_i \over \rho_i} -{V_i \over \rho_i} {d\over dx} \delta \rho_i + \xi {\rho_n \over \rho_i} {\delta \rho_n \over \rho_n} - {d v_i \over dx} = \rmi \omega {\delta \rho_i \over \rho_i} ,\\
 -{c_s^2 \over \rho_i} {d\over dx} \delta \rho_i - \gamma \delta \rho_n V_d  - V_i {d\over dx} v_i + \gamma \rho_n (v_n -v_i) -{V_{A,i}^2 \over B} {d\over dx} \delta B=\rmi \omega v_i,\\
 -{d \over dx} v_i - {V_i \over B} {d \over dx} \delta B = \rmi \omega {\delta B \over B},\\
-{V_n\over \rho_n} {d \over dx} \delta \rho_n - {d \over dx} v_n = \rmi \omega {\delta \rho_n \over \rho_n}, \\
\gamma V_d \delta \rho_i  -{c_s^2 \over \rho_n} {d\over dx} \delta \rho_n + \gamma \rho_i (v_i - v_n) - V_n {d \over dx} v_n = \rmi \omega v_n.
\end{eqnarray}
}

%Let $U \equiv (\delta \rho_i,v_i,\delta B,\delta \rho_n, v_n)^T$. The above equations can then be written as 
\begin{equation}
C U = \rmi \omega U,\label{eq:disp_matrix}
\end{equation}
where
\begin{eqnarray}
C=\left[\arraycolsep=6pt
\begin{array}{ccccc}
-\rmi k V_i-2\beta \rho_i& -\rmi k \rho_i  & 0 & \xi_\mathrm{CR} & 0 \\
-\rmi k \frac{c^2_s}{\rho_i} & -\rmi k V_i -\gamma \rho_n & -\rmi k \frac{V^2_{A,i}}{B}  & -\gamma V_d & \gamma \rho_n \\
0 & -\rmi k B & -\rmi k V_i & 0 & 0 \\
0 & 0 & 0 & -\rmi k V_n  & -\rmi k \rho_n  \\
\gamma V_d & \gamma \rho_i & 0 & -\rmi k \frac{c_s^2}{\rho_n}  & -\rmi k V_n-\gamma \rho_i 
\end{array}
\right].
\end{eqnarray}
Hence, we can solve the above equations as an eigenvalue problem with $\omega$ being the eigenvalue and $U$ being the eigenfunctions. The goal is to identify a maximum-growth mode associated with the drag instability within the C-shock.

\subsection{Fiducial Model}
\label{sec:model}

We adopt the 1D steady-state C-shock profile shown in Figure~3 of \citet{CO12} as the background state (in the shock frame) of our fiducial model.
The pre-shock parameters are
%$n_0=500$ cm$^{-3}$, $v_0=5$ km/s, $B_0=5 \mu$G, and $\chi_{i0}=10$ (see their Figure 2).  
$n_0=500$ cm$^{-3}$ (neutral number density), $V_{i,0}=V_{n,0}=v_0=5$ km/s (shock velocity), $B_0=10 \mu$G, and ionization fraction coefficient $\chi_{i0}=10$. Here, the parameter $\chi_{i0}$ is defined in the expression $n_i=10^{-6} \chi_{i0}n_n^{1/2}$ of \citet{CO12} assuming ionization-recombination equilibrium, and is therefore given by $\chi_{i0} \equiv 10^6 \sqrt{\xi_\mathrm{CR} (m_n/m_i)/(\beta m_i)}$. 
We thus adopted $\beta \approx 10^{-7}$~cm$^3$~s$^{-1}/m_i$ and $\xi_\mathrm{CR} \approx 10^{-17}$~s$^{-1}$~($m_i/m_n$) in this study \citep[see, e.g.,][]{Shu,Tielens05}, where $m_n=2.3\times$ and $m_i=30 \times$ the hydrogen mass are considered.
Indeed, $\chi_{i0}=10$ falls in the typical range of $\chi_{i0}$ observed in star-forming regions ($\sim 1-20$; see, e.g.,~\citealt{McKee10}).
The fiducial model is selected such that it can be reproduced easily by comparison with the results of \citet{CO12}.

\begin{figure}
\plottwo{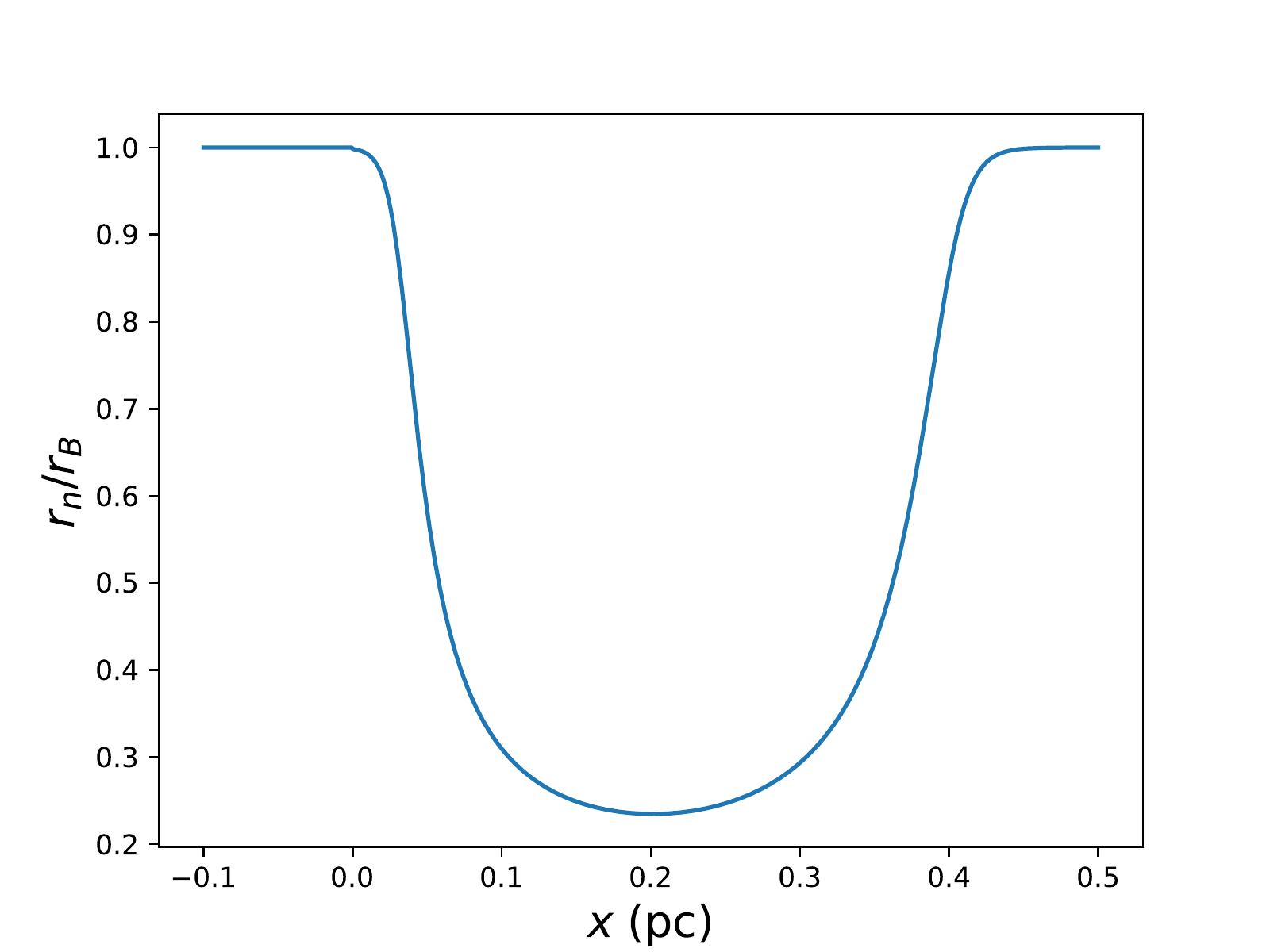}{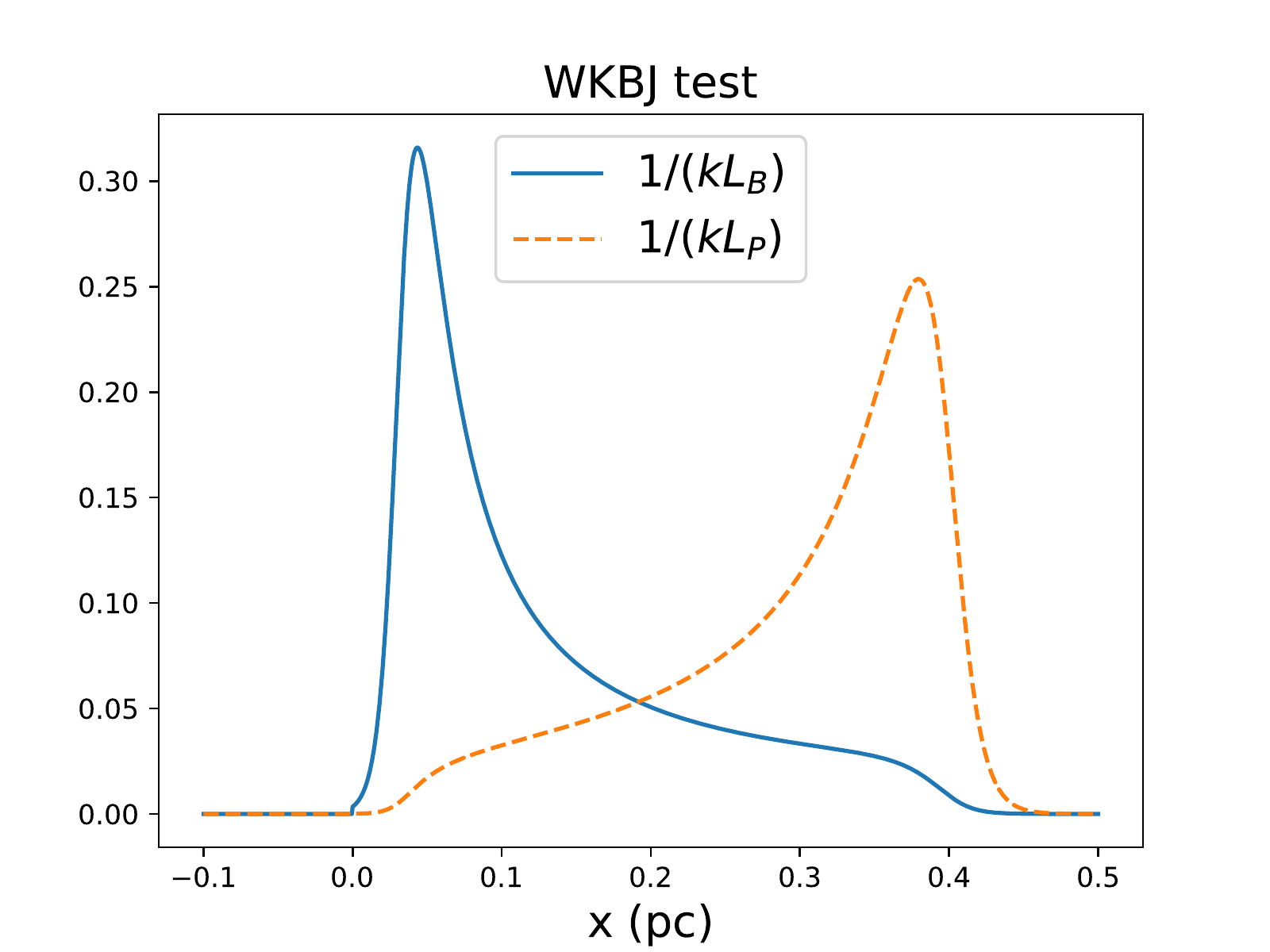}
\caption{The $r_n/r_B$ ratio of the background state throughout the C-shock (left panel) and a test for the WKBJ approximation when $k=1/0.015$ pc$^{-1}$ (right panel). $L_p \equiv |d\ln p/dx|^{-1}$ and $L_B \equiv |d \ln B/dx|^{-1}$ are the scale heights for the gas pressure and magnetic field of the background states, respectively.}
\label{fig1}
\end{figure}

Without the loss of generality, we adopt a constant wavenumber ($k$) of $1/0.015$ pc$^{-1}$ to keep the fiducial model as simple as possible. Figure~\ref{fig1} displays the $r_n/r_B$ ratio of the background state (left panel) and the validity of the WKBJ approximation (right panel). The C-shock transition begins from $x=0$ pc and ends at approximately $x=0.4$ pc. We obtained the same U-shape for the $r_n/r_B$ ratio as that obtained by \citet{CO12} in their Figure 3. The U-shaped profile is a notable feature of the C-shock model, as demonstrated by \citet{CO12}. The ratio $r_n/r_B=1$ in the pre- ($x<0$ pc) and post-shock ($x>0.4$ pc) regions where no background gradients are present. Throughout the C-shock, $1/(kL_B)$ and $1/(kL_p)$ are considerably smaller than 1; thus, the WKBJ approximation is justified. The marginally high value of $1/(kL_B)$ and $1/(kL_p)$ around $x=0$ pc and $x=0.4$ pc, respectively, is caused by the initial compression of the ions and thus the magnetic fields, followed by a delayed compression of the neutrals by means of the ion-neutral drag as the gas flows downstream across the steady C-shock.

%We compute the growth rate within the C-shock based on the linear WKBJ result from \citet{Gu04} with the wavenumber $k=1/0.01$ pc$^{-1}$. 
%The results are shown in Figure~\ref{fig1}. The ratio $r_n/r_B$ is the same as Figure 2 in \citep{CO12}. 

%With this background state, the right panel of Figure~\ref{fig1} shows that within the C-shock (i.e. $ x \approx 0.05-0.25$ pc), 
\subsection{Dispersion Relations}
\label{sec:dispersion}

Before solving the eigenvalue problem in Equation(\ref{eq:disp_matrix}), we analyze Equation(\ref{eq:disp_matrix}) in terms of a couple of simplified dispersion relations, which will provide the basic and clear physics to better understand the exact solutions obtained using the eigenvalue approach. Table~\ref{tab:sym} summaries the meanings of the main symbols that we are going to use in the linear analysis.

\begin{table*}
\begin{center}
  \caption{ Summary of the main symbols adopted in the linear analysis. }
	\label{tab:sym}
\begin{tabular}{ll } 
		\hline
		Symbol \& definition & meaning \\
		\hline
%		$L_p \equiv |d\ln p/dx|^{-1}$ & scale height for the gas pressure $p$ within C shocks\\ 
%		$L_B \equiv |d \ln B/dx|^{-1}$ & scale height for the magnetic field $B$ within C shocks\\
%		\hline
		$\Gamma \equiv \rmi \omega + \rmi k V_n$ & rate of a mode in the comoving frame of the neutrals\\
		$\Gamma_{GLV}$ & rate of the unstable/decaying mode derived by \citet{Gu04}\\
		$\Gamma_{re} \equiv 2\beta\rho_i$ & the recombination rate\\
		$\Gamma_{grav} \equiv \sqrt{G\rho_n}$ & rate of the gravitational instability\\
		$\Gamma_{th} \equiv k c_s$ & sound-crossing rate over one wavelength\\
		$\Gamma_{kV_d} \equiv k|V_d|$ & the ion-neutral drift rate across a distance of wavelength\\
		$\Gamma_{alf,i} \equiv k V_{A,i}$ & the speed of the Alfv{\'e}n wave in the ions crossing one wavelength \\
		$\Gamma_{alf,n} \equiv k V_{A,n}$ & the speed of the Alfv{\'e}n wave in the neutrals crossing one wavelength \\
		$\Gamma_{ambi}\equiv k^2 D_{ambi}$ & ambipolar diffusion rate \\
		$\gamma_n \equiv \gamma \rho_n$ &  the ion collision rate with the neutrals\\
		$\gamma_i \equiv \gamma \rho_i$ &  the neutral collision rate with the ions\\
		$\Gamma_{grow}$ & the growth rate of an unstable mode\\
		$\omega_{wave,n}$ & the wave frequency of an unstable mode in the comoving frame of the neutrals\\
		$\omega_{wave}\equiv {\rm Re}[\omega]$ & the wave frequency of a mode\\
		\hline
		$D_{ambi} \equiv V_{A,n}^2/\gamma \rho_i$ & ambipolar diffusion coefficient \\
		$\gamma \approx 3.5 \times 10^{13}$~cm$^3$~s$^{-1}$~g$^{-1}$ & the drag force coefficient \citep{Draine} \\
		\hline
	\end{tabular}
	\end{center}
\end{table*}

The analysis begins  with a brief review of the drag instability. \citet{Gu04} indicated that when the drift velocity $V_d$ is sufficiently high, a type of overstability, called the drag instability, can occur. Specifically, for the definite occurrence of the instability,
the rate of the mode $\Gamma \equiv \rmi \omega + \rmi k V_n$ observed in the comoving frame of the neutrals is considerably lower than both the recombination rate $\Gamma_{re} \equiv 2 \beta \rho_i$ and the ion-neutral drift rate across a distance of wavelength (i.e., $\Gamma_{kV_d} \equiv k|V_d|$), whereas it is considerably higher than the neutral collision rate with the ions (i.e., $\gamma_i \equiv \gamma \rho_i$), the sound-crossing rate over one wavelength (i.e., $\Gamma_{th} \equiv k c_s$), and the rate of the gravitational instability (i.e., $\Gamma_{grav} \equiv \sqrt{G\rho_n}$). Although self-gravity is not included in our equations, $\Gamma_{grav}$ is still estimated to evaluate its significance.
%and the wave crossing rate of the instability through the shock width $\Gamma_{crossing}\equiv v_{ph}/L_{shock} \approx (3\, {\rm km/s})/(0.2\,{\rm pc})$.  
When the aforementioned conditions are satisfied, the linearized equation (Equation~\ref{eq:disp_matrix}) is substantially reduced to \citep[see][]{Gu04}
\begin{eqnarray}
\Gamma {\delta \rho_n \over \rho_n}=\rmi k \delta v_n, \label{eq:n_cont}\\
\Gamma \delta v_n \approx \gamma \rho_i V_d {\delta \rho_i \over \rho_i},\label{eq:n_drag}\\
2 \beta \rho_i {\delta \rho_i \over \rho_i} \approx \xi_\mathrm{CR} {\rho_n \over \rho_i} {\delta \rho_n \over \rho_n}.\label{eq:ion_eq}
\end{eqnarray}
Along with the background states, the above equations leads to the dispersion relation,
\begin{equation}
\Gamma_{GLV} \approx \pm {(1+\rmi)\over 2} \sqrt{\Gamma_{kV_d} \gamma_i }=\pm  {(1+\rmi)\over 2}  \sqrt{k \over L_B} V_{A,n}. \label{eq:disp_Gu}
\end{equation}
In the above dispersion relation, the positive and negative parts correspond to the growing and decaying waves, respectively. The subscript GLV is added to $\Gamma$ to indicate the growth/damping rate (Re[$\Gamma$]) and wave frequency (Im[$\Gamma$]) of the unstable/decaying mode derived by \citet{Gu04}, which is compared to the eigenvalue of the growing mode in \S\ref{sec:eigen} later in the paper. The growth/damping rate is lower than the speed of the Alfv\'en wave in the neutrals crossing one wavelength $\Gamma_{alf,n}\equiv kV_{A,n}$, which indicates that the unstable/decaying mode is slower than the magneto-acoustic mode of the bulk fluid.
% and thus is almost ``incompressible" in this regard. 
It is also evident that the above dispersion relation appears very different from that for the magneto-acoustic mode, and therefore the density perturbation is not caused by magneto-acoustic oscillations\footnote{In this regard, the unstable mode is ``incompressible."}. The reason why the 1D overdensity/underdensity occurs in the bulk of the fluid (i.e., the neutrals) is that the neutrals experience high drag due to the density clump of the ions (Equation \ref{eq:n_drag}), which is also the density clump of the neutrals due to  the rapid ionization equilibrium (Equation \ref{eq:ion_eq}).

\begin{figure}
\plotone{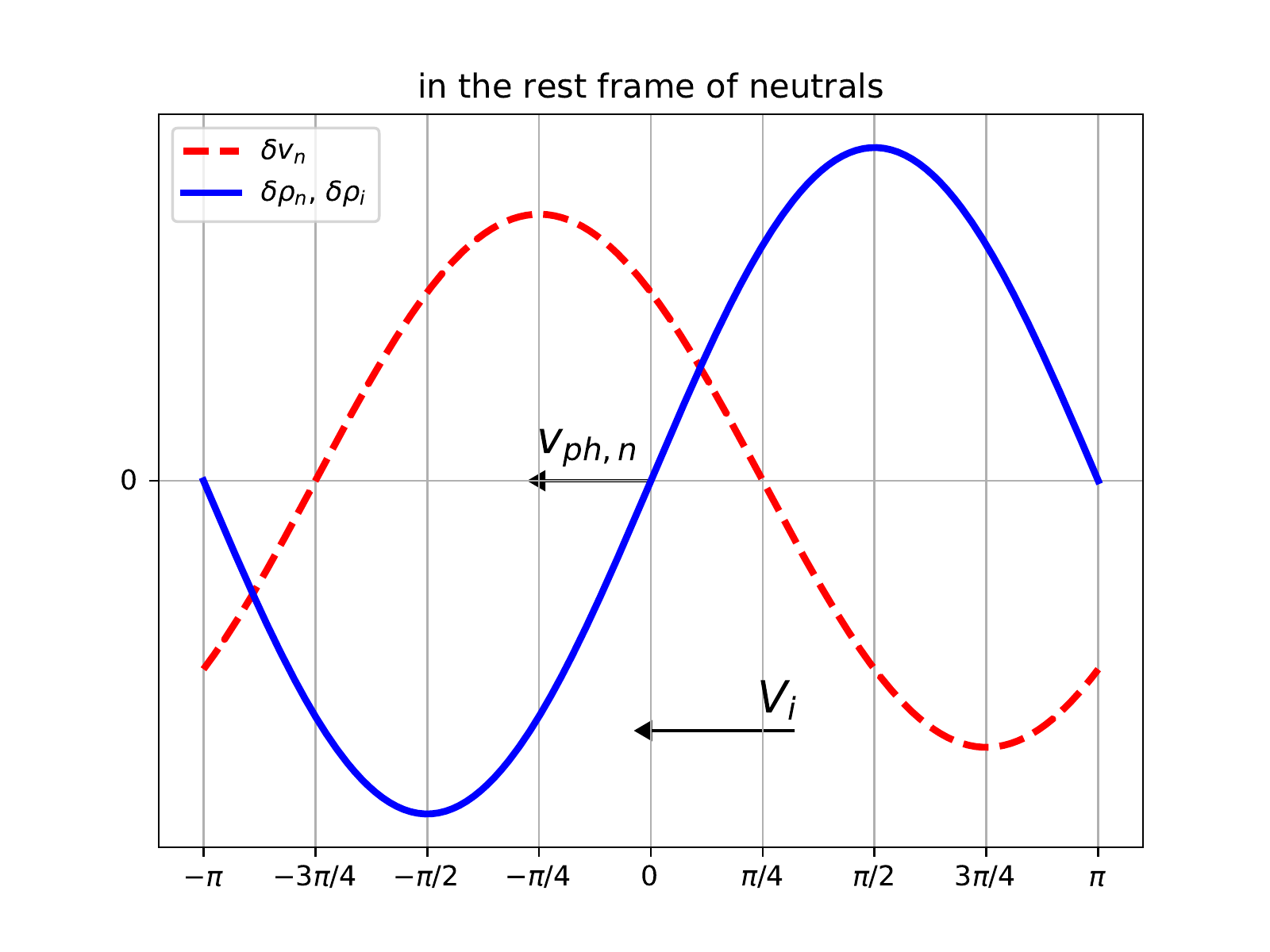}
\caption{Phase diagram of the drag instability in the rest frame of the neutrals. The profiles of $\delta v_n$ and $\delta \rho_n$ as a function of $x$ are displayed, and the directions of $V_i$ and $v_{ph,n}$ are indicated. The amplitudes of the perturbations are shown on arbitrary scales.}
\label{fig:phase_diff}
\end{figure}

By substituting the positive part of Equation \ref{eq:disp_Gu} (for an unstable wave) into Equation~(\ref{eq:n_cont}), we have the relation
$\delta v_n \propto \delta \rho_n \exp(\rmi 3\pi/4)$ and obtain the phase velocity in the comoving frame of the neutrals given by $v_{ph,n}=-{\rm Im}[|\Gamma_{GLV}|]/k<0$. These imply that $v_n$ leads $\delta \rho_n$ by a phase of $3\pi/4$, and the unstable wave travels upstream in the rest frame of the neutrals. As explained in \citet{Gu04}. this specific phase difference between $\delta v_n$ and $\delta \rho_n$ in the rest frame of neutrals is the physical origin of the drag instability. 
Figure~\ref{fig:phase_diff} illustrates how the instability occurs. The ion and neutral density perturbations $\delta \rho_i$ and $\delta \rho_n$ are in phase due to the ionization equilibrium. In the rest frame of the neutrals, the wave and $V_i$ propagate upstream (to the left), and $\delta v_n$ leads $\delta \rho_n$ by a phase difference $3\pi/4$.  Consequently, the peak of the density perturbation $\delta \rho_n$ at $x=\pi/2$ continues to increase due to the converging velocity field (i.e., $d \delta v_n/dx <0$), while the trough of the density perturbation at $x=-\pi/2$ continues to decrease due to the diverging velocity field (i.e., $d \delta v_n/dx >0$),
%the top 75\% of the overdensity (i.e. the positive $\delta \rho_n$ from phase 0 to $3\pi/4$ in the figure) is compressed by converging velocity fields ($d \delta v_n/dx <0$), 
thereby leading to further growth of the perturbations. We refer to the study of \citet{Gu04} for more detailed descriptions\footnote{In the study of \citet{Gu04}, $V_d>0$ and thus $\delta v_n$ leads $\delta \rho_n$ by a phase of $\pi/4$ instead of $3\pi/4$ in the comoving frame of the neutrals. Nevertheless, the image for the drag instability is the same.}.
Note that the unstable wave in fact propagates downstream in the shock frame at the phase velocity given by $v_{ph,n}+V_n$. Namely,  the fast streaming motion of the background flow brings the growing wave downstream through the shock.

We now examine the dispersion relations outside a C-shock where $V_i=V_n$. Thus, $V_d=0$ and $\Gamma_i \equiv \rmi \omega+ \rmi k V_i=\Gamma$ in both the pre- and post-shock regions. Hence, the dispersion relation derived from Equation(\ref{eq:disp_matrix}) reads \citep[cf.][]{Gu04}
\begin{equation}
%(\Gamma+2\beta \rho_i) [(\Gamma_{alf,i}^2 + \Gamma^2)[\Gamma_{th}^2 + \Gamma(\Gamma+\gamma_i)]+\gamma_n \Gamma (\Gamma_{th}^2+ \Gamma^2) ]=0,\label{eq:disp_orig}
\Gamma_{th}^4\Gamma + \Gamma \Gamma_{re} [ \Gamma_{alf,i}^2( \Gamma+ \gamma_i) + \Gamma^2 \gamma_n ] + \Gamma_{th}^2 \left[ \Gamma_{alf,i}^2 \Gamma_{re} + \Gamma (2\Gamma +\gamma_i) {\Gamma_{re}\over 2} + \Gamma \Gamma_{re} \gamma_n \right]=0,  \label{eq:disp_orig}
\end{equation}
where $\Gamma_{alf,i} \equiv k V_{A,i}$ and $\gamma_n \equiv \gamma \rho_n$. 
%It is obvious that there is a mode with the damping rate of $\Gamma=-2\beta \rho_i$ due to recombination. Furthermore, 
The rate $\Gamma_{th}$ usually has a low value. It may be even lower than $\gamma_i$ in the above dispersion relation, which sometimes occurs in the post-shock region where the neutral density is compressed and the ion density is subsequently enhanced by ionization. %Excluding the decaying mode due to recombination and 
Ignoring the thermal terms associated with $\Gamma_{th}$ at this moment, Equation(\ref{eq:disp_orig}) can be reduced as follows
\begin{equation}
(\Gamma + \Gamma_{re})[(\Gamma_{alf,i}^2 + \Gamma^2)(\Gamma+ \gamma_i ) + \gamma_n \Gamma^2] =0.
\end{equation}
The first term of the above equation represents a decaying mode with a damping rate of $-\Gamma_{re}$.  In the second term of the above equation,
we have recovered the well-known dispersion relation for 1D linear Alfv\'en waves in a weakly ionized plasma \citep{KP69}\footnote{\citet{KP69} used the perturbations $\propto \exp(\rmi kx-\rmi \omega t)$, whereas we use the perturbations $\propto \exp(\rmi kx+\rmi \omega t)$ and consider a background flow that produces the Doppler-shifted frequency $kV_n$.}. Two branches of this mode can exist depending on the strength of the coupling between the ions and the neutrals (i.e., the strong and weak-coupling branches). In the strong-coupling branch, $\Gamma_{alf,i} \ll \gamma_n$; thus, the dispersion relation can be further reduced as follows \citep[cf.][]{KP69,McKee10}
\begin{equation}
%\omega + k V_n = \pm \left( k^2V_{A,i}^2 -{\gamma^2 \rho_n^2 \over 4} \right)^{1/2}+ \rmi {\gamma \rho_n \over 2}.
\omega + k V_n =\pm \left(  \Gamma_{alf,i}^2 {\rho_i \over \rho_n} - {\Gamma_{alf,i}^4 \over 4 \gamma_{n}^2} \right)^{1/2} + \rmi { \Gamma_{alf,i}^2 \over 2 \gamma_{n}} \approx \rmi \left[ {\Gamma_{ambi}\over 2}  \pm \left( {\Gamma_{ambi}\over 2}- \gamma_i \right) \right], \label{eq:disp1}
\end{equation}
where $\Gamma_{ambi}$ is the ambipolar diffusion rate equal to $k^2 D_{ambi}$, with the ambipolar diffusion coefficient $D_{ambi} \equiv V_{A,n}^2/\gamma \rho_i$. We have assumed that $\Gamma_{ambi} \gg \Gamma_{alf,n}$ to expand the expression of the square root to further simplify the final result on the right-hand side of Equation(\ref{eq:disp1}). On the other hand, the dispersion relation is obtained for the weak-coupling regime  \citep{KP69,McKee10}
\begin{equation}
\omega + k V_n =\pm \left(   \Gamma_{alf,i}^2 - {\gamma_n^2 \over 4} \right)^{1/2} + \rmi {\gamma_n \over 2} \approx \rmi \left[ {\gamma_n \over 2} \pm \left( {\gamma_n \over 2} - \Gamma_{ambi} \right) \right],\label{eq:disp2}
\end{equation}
 where we have assumed that $\gamma_n^2 \gg \Gamma_{ambi}^2$, which is equivalent to $\gamma_n \gg \Gamma_{alf,i}$, to expand the expression of the square root to derive the right-hand side of Equation(\ref{eq:disp2}). Thus, Equations(\ref{eq:disp1}) \& (\ref{eq:disp2}) indicate that no waves but decaying modes exist
 in the frame comoving with the flow. The damping rates in the strong-coupling branch are $\sim \Gamma_{ambi}$ and $\gamma_i$ and the damping rates in the weak-coupling branch are $\sim \gamma_n$ and $\Gamma_{ambi}$. 
 
%General speaking, the sign of the expression inside the square root in the above two dispersion relations depends on $k$. In our fiducial model with $k=1/0.015$ pc$^{-1}$, the inequality $\gamma_n \gg \Gamma_{ambi} \gg \Gamma_{alf,n}$ holds and thus the expressions are negative to justify the approximations made in deriving the dispersion relations eq(\ref{eq:disp1}) \& eq(\ref{eq:disp2}). Therefore, in the frame comoving with the flow, no waves but decaying modes exist. From equations(\ref{eq:disp1}) \& (\ref{eq:disp2}), the damping rates in the strong coupling branch are $\sim \Gamma_{ambi}$ and $\gamma_i$. The damping rates in the weak coupling branch are $\sim \gamma_n$ and $\Gamma_{ambi}$. 

Finally, there exist slow decaying modes associated with the weak thermal effect that we have ignored so far.  When we consider that $\Gamma \ll \Gamma_{alf,i}$, $\gamma_i$ and because $\Gamma_{th} \ll \Gamma_{ambi}$, Equation(\ref{eq:disp_orig}) can be reduced to $\Gamma_{th}^2+\Gamma^2+\Gamma \gamma_i \approx 0$, which has two solutions: $\Gamma \approx (-\gamma_i \pm \sqrt{\gamma_i - 4 \Gamma_{th}^2})/2$.
If $\gamma_i^2 \gg 4 \Gamma_{th}^2$, the reduced dispersion relation yields the following two decaying modes with no waves in the comoving frame of the flow:
\begin{equation}
\omega + kV_n \approx  \rmi {\Gamma_{th}^2 \over \gamma_i} ,\ \rmi \gamma_i.\label{eq:post_decay1}
\end{equation}
However, if $\gamma_i^2 \ll 4 \Gamma_{th}^2$, the following two decaying wave modes exist in the comoving frame of the flow:
\begin{equation}
\omega + kV_n \approx  \rmi {\gamma_i \over 2} \pm \Gamma_{th}.\label{eq:post_decay2}
\end{equation}
%$\Gamma_{alf,i}^2\Gamma_{th}^2 + \Gamma_{alf,i}^2 \Gamma \gamma_i \approx 0$, which leads to a mode with the decay rate $\Gamma \approx -\Gamma_{th}^2/\gamma_i$.

In summary, the dispersion relation in the post-shock region (Equation(\ref{eq:disp_orig})) indicates the presence of  five decaying modes with the damping rates of $\gamma_n$, $2\beta \rho_i$, $\Gamma_{ambi}$, $\gamma_i$, and $\Gamma_{th}^2/\gamma_i$ when $\Gamma_{th} < \gamma_i$ or $\gamma_n$, $2\beta \rho_i$, $\Gamma_{ambi}$, $\gamma_i/2$, and $\gamma_i/2$ (the same as for the two slowest modes) when $\Gamma_{th} > \gamma_i$.
%In our fiducial model, it can be shown that this decay rate is much larger than the flow crossing rate through one wavelength $k V_n$ outside the shock and also larger than the growth rate in the shock, probably implying a quick decay once the growing wave travels out of the shock.

%This gives rise of a wave that travels at the ion Alfven speed in the flow frame and damps at the rate $\ \propto \gamma \rho_n$.  
\begin{figure}
\plottwo{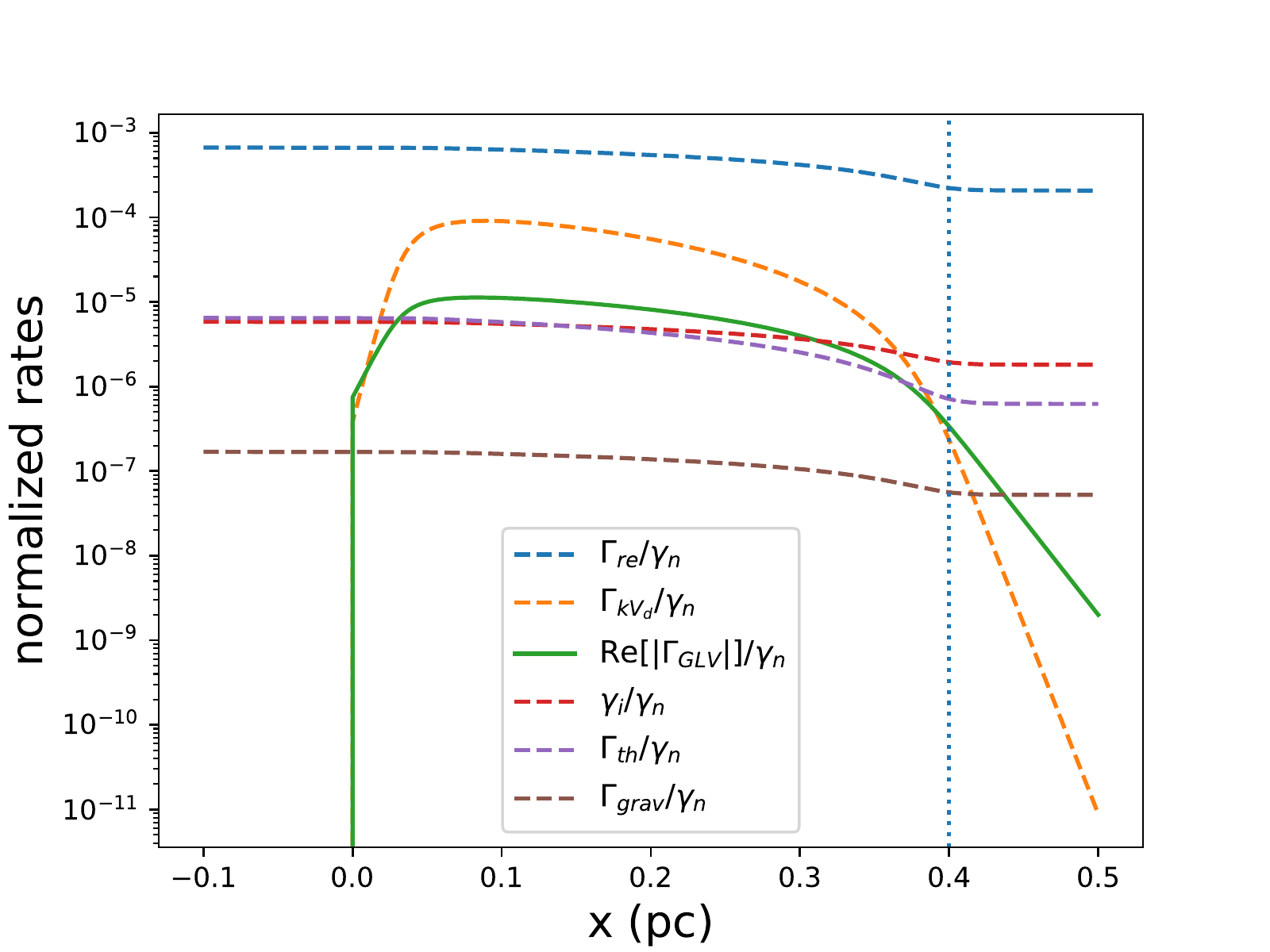}{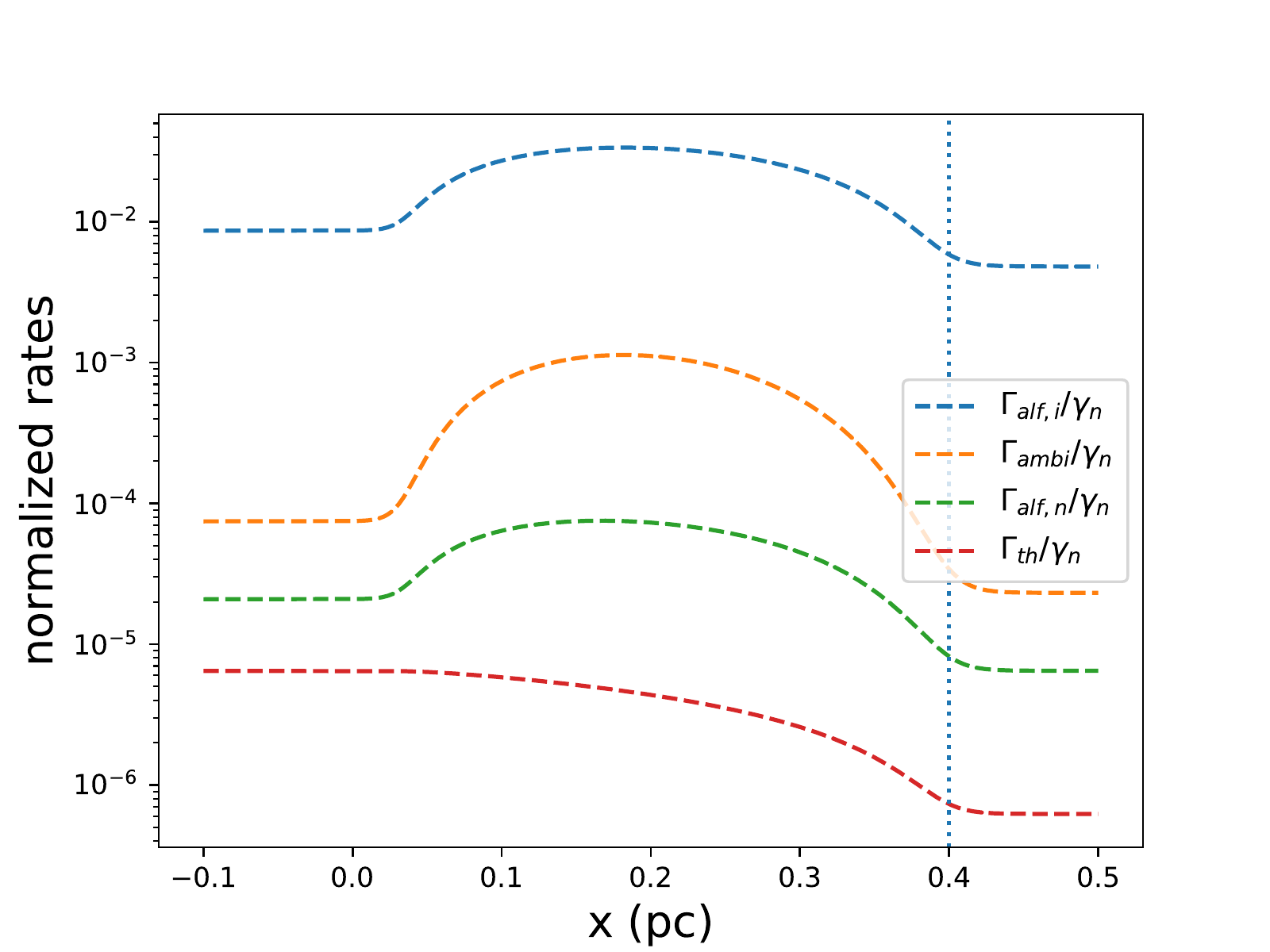}
\caption{Comparison of various rates relevant to the assumptions for the derivation of dispersion relations, normalized by $\gamma_n$ throughout the C-shock  in our fiducial model (left panel).  Additional normalized rates relevant to the post-shock regions are also plotted and compared (right panel). The domain to the right of the vertical dotted line represents the post-shock region.}
\label{fig2}
\end{figure}

Figure~\ref{fig2} illustrates the suitability of the assumptions made in our fiducial C-shock model, where $k$ is set to $1/0.015$ pc$^{-1}$; the assumptions were made to derive the dispersion relations inside and outside the C-shock. The left panel of Figure~\ref{fig2} indicates that the ion-neutral drift rate across one wavelength, $\Gamma_{kV_d}$, becomes high within the C-shock (i.e. between $x=0$ and $x\approx 0.4$ pc) because of the high drift velocity $V_d$ that is caused by the shock compression. Consequently, $\Gamma_{kV_d}$ is higher than $\gamma_i$ but still lower than $\Gamma_{re}$ and $\gamma_n$ inside the C-shock. The drag instability thus occurs inside the shock according to the dispersion relation presented in Equation~(\ref{eq:disp_Gu}), with the growth rate Re[$\Gamma_{GLV}$] being higher than $\gamma_i$ and $\Gamma_{th}$. Although we plot $\Gamma_{GLV}$ beyond $x\approx 0.4$ pc in the post-shock region, the instability is expected to disappear because $\Gamma_{kV_d}$ decreases quickly outside the C-shock.
We also plot the rate of gravitational instability $\Gamma_{grav}$, which is considerably lower than the other rates and can be reasonably neglected in the linear analysis.

The right panel of Figure~\ref{fig2} indicates that  $\gamma_n \gg \Gamma_{alf,i} \gg \Gamma_{ambi} \gg \Gamma_{alf,n}$, which ensures the presence of the decaying modes described by the dispersion relations with $V_d=0$, i.e., Equations~(\ref{eq:disp1}) and (\ref{eq:disp2}), in the post-shock region. We also see from the figure that 
 $\Gamma_{th}$ is the lowest rate of the rates of interest. It can be ignored except when $\Gamma \lesssim \Gamma_{th}$, which results in the presence of two modes of the lowest decaying rate associated with $\gamma_i$ and $\Gamma_{th}$ in the post-shock region, as described by Equations~(\ref{eq:post_decay1}) \& (\ref{eq:post_decay2}).  In the fiduical model, the left panel of Figure~\ref{fig2} shows that $\gamma_i > \Gamma_{th}$ in the post-shock region. Hence, the decaying mode is expected to follow the dispersion relation described by Equation~(\ref{eq:post_decay1}) more closely (see the next subsection).
 
\subsection{Exact Solutions Obtained from the Eigenvalue Problem}
\label{sec:eigen}

\begin{figure}
\plottwo{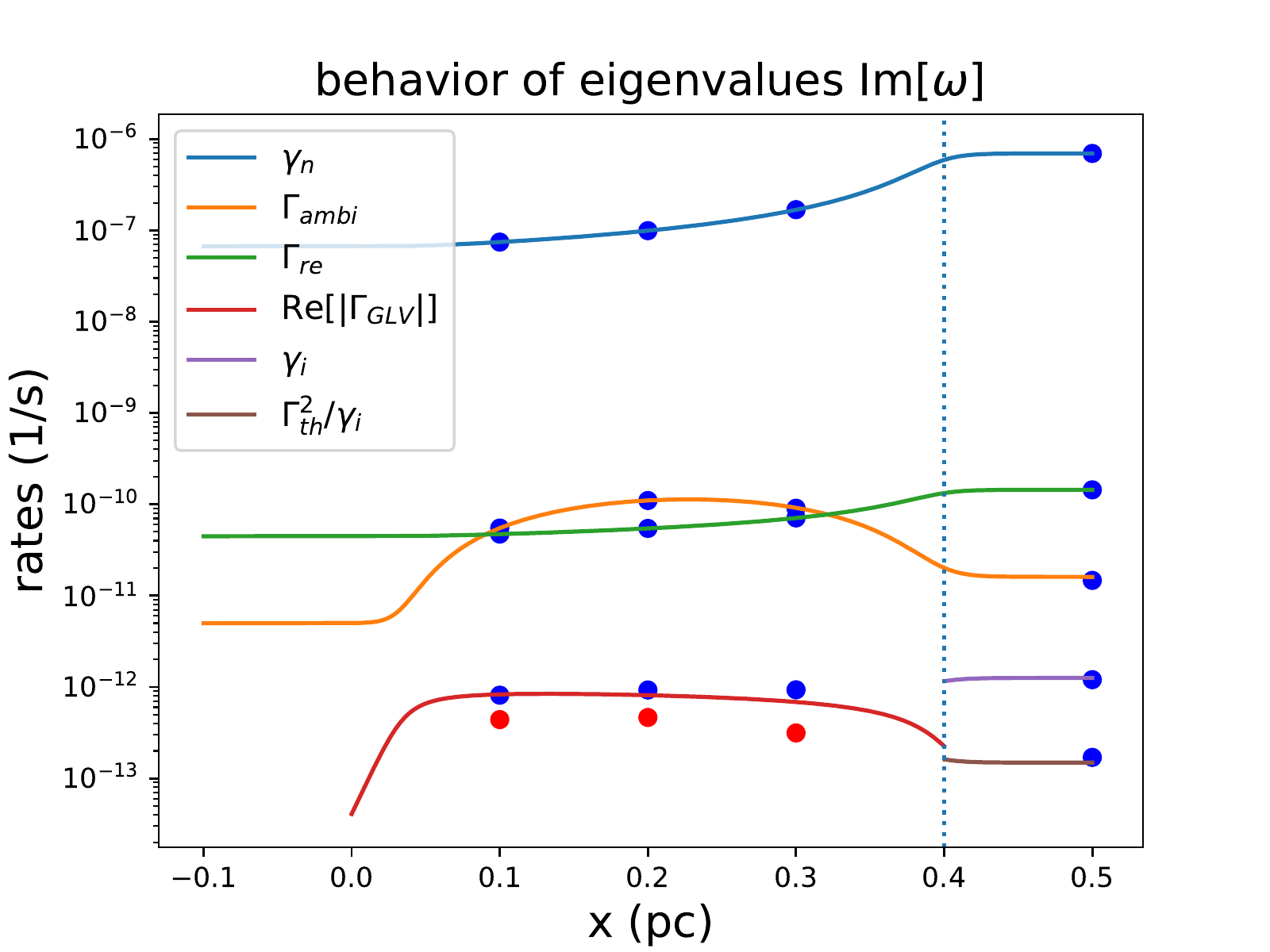}{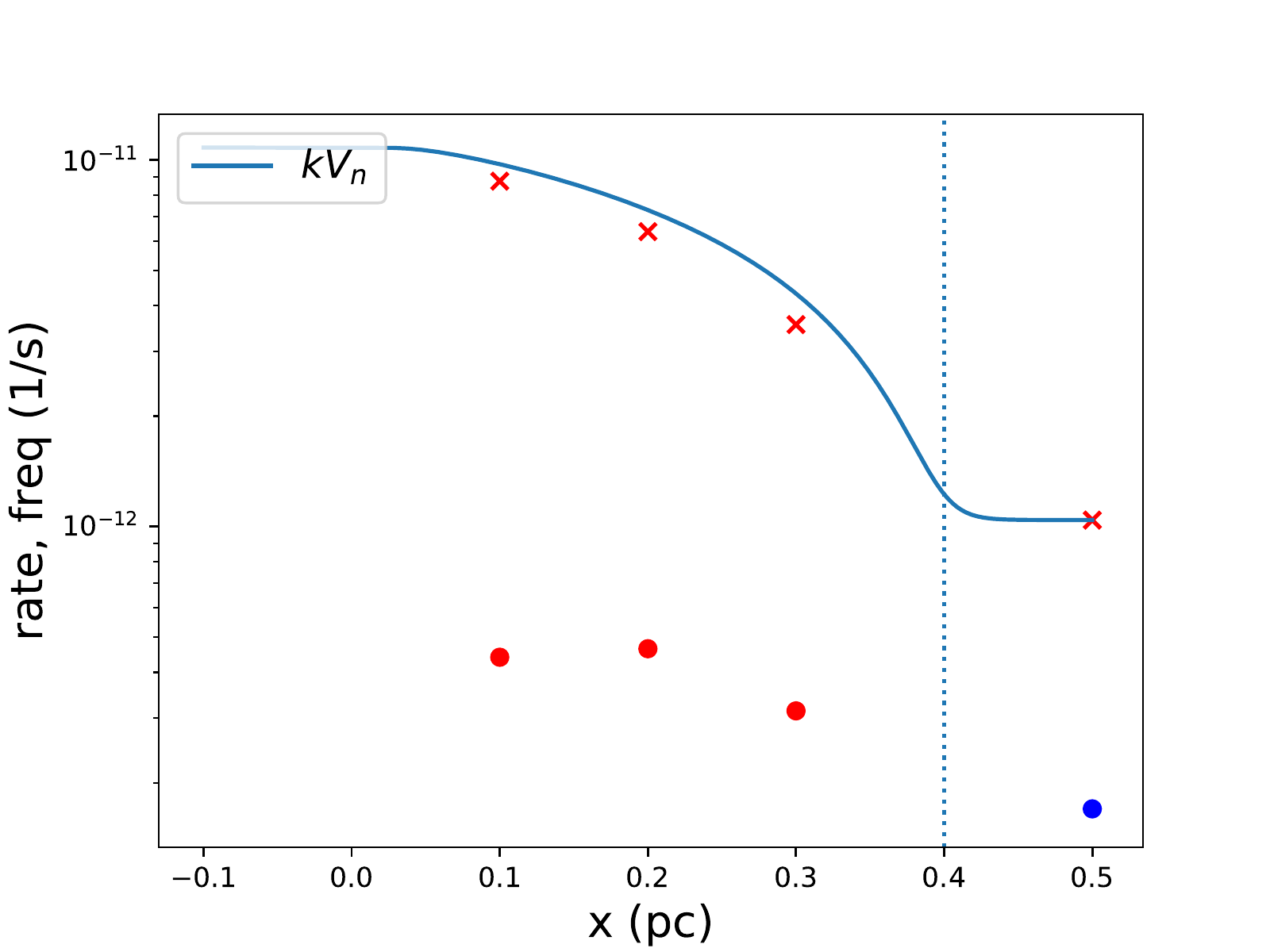}
\caption{Properties of the eigenvalues in the shock and post-shock regions in the fiduical model. The left panel shows 5 eigenvalues Im[$|\omega|$] at $x=0.1$, 0.2, 0.3, and 0.5 pc in terms of blue (Im[$\omega]>0$ indicative of a damping rate) and red (Im[$\omega]<0$ indicative of a growth rate) dots. The curves for various rates relevant to the eigenvalue $\omega$ are overplotted for comparison. The right panel shows the wave frequencies Re[$-\omega$] (red crosses) associated with the modes with the smallest Im[$|\omega|$] (colored dots), i.e., the unstable modes inside the shock and the mode with the slowest damping rate in the post-shock region. The curve for $kV_n$ is also plotted for comparison. The domain to the right of the vertical dotted line ($x\approx 0.4$~pc) in the two panels is the post-shock region.}
\label{fig:eigenvalue}
\end{figure}

\begin{figure}
\plottwo{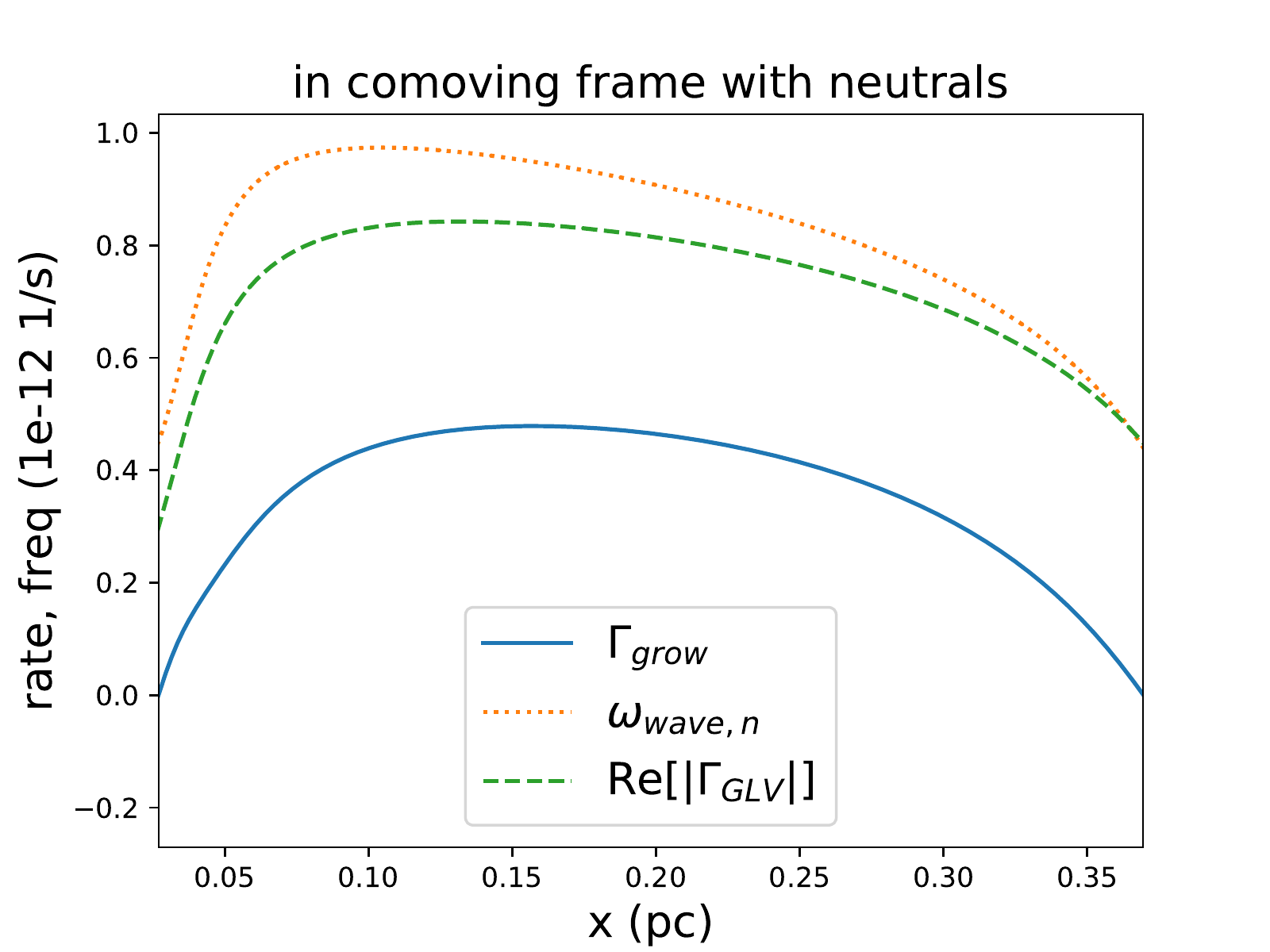}{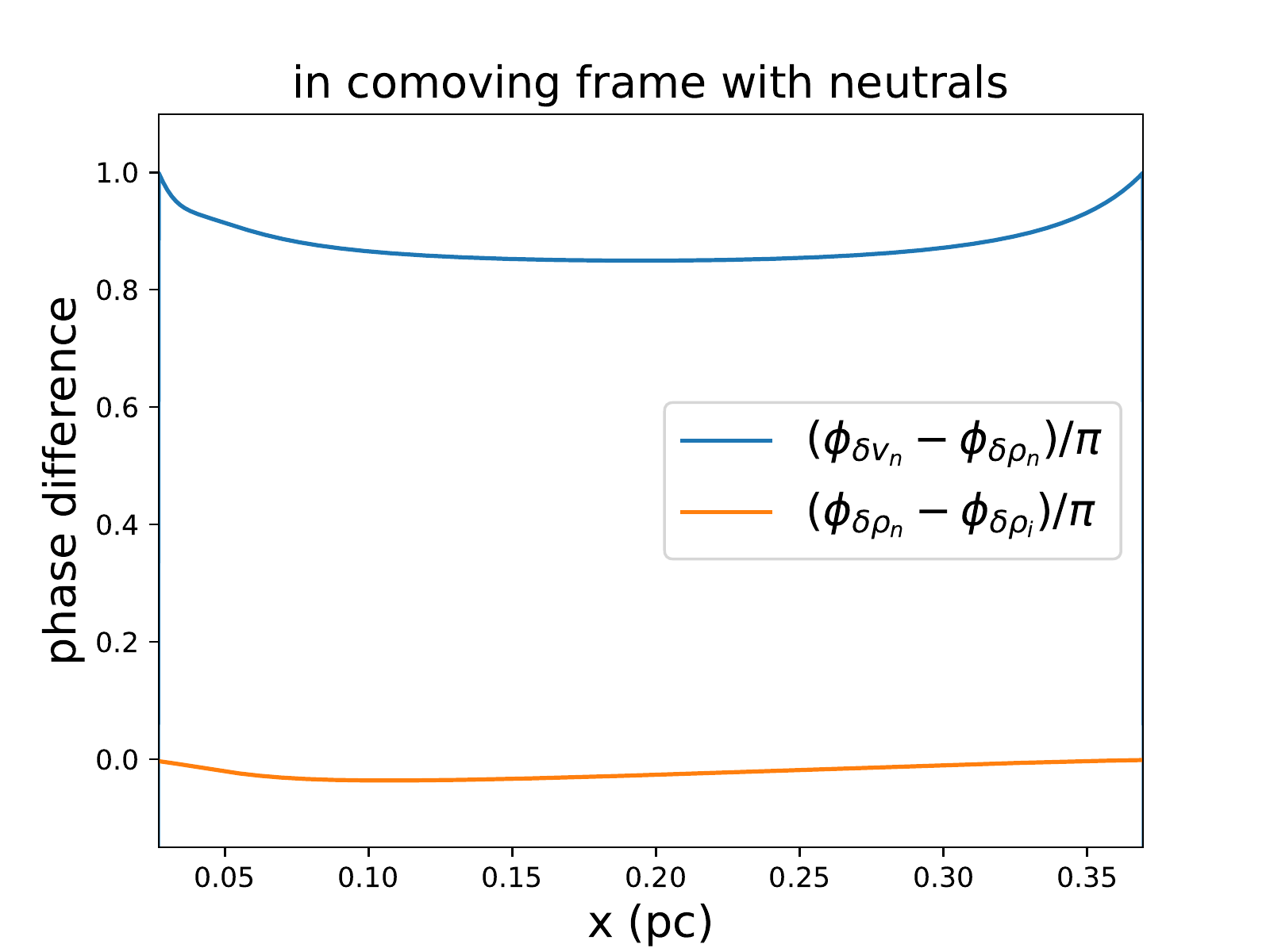}
\caption{The growth rate $\Gamma_{grow}$ and wave frequency $\omega_{wave,n}$ of the unstable mode seen by the neutrals are plotted in comparison with the growth rate Re[$\Gamma_{GLV}$] estimated by \citet{Gu04} (left panel).
The phase difference between the velocity and density perturbations of the neutrals ($\phi_{\delta v_n}-\phi_{\delta \rho_n}$) and the phase difference between the density perturbation of the neutrals and the ions ($\phi_{\delta \rho_n} -\phi_{\delta \rho_i}$) are also displayed within the C-shock (right panel).}
\label{fig3}
\end{figure}

%After identifying the modes and determining their underlying physics through the simple dispersion relations, we study the exact solutions of the entire set of linearized equations in Equation(\ref{eq:disp_matrix}). Table~\ref{tab:eigen} lists the eigenvalues inside the C-shock at $x=0.1$ and 0.3 pc as well as the eigenvalues in the post-shock region at $x=0.5$ pc. Of the five eigenmodes at $x=0.1$ and 0.3 pc, only one unstable wave mode exists inside the C-shock (i.e. positive Im[$-\omega_5$] but negative Im[$-\omega_1$]-Im[$-\omega_4$]). By contrast, all the decaying wave modes exist at $x=0.5$ pc in the post-shock region (i.e. negative Im[$-\omega_1$]-Im[$-\omega_5$]). To understand the properties of these eigenvalues, 
%the rates relevant to the different decaying modes at the same $x$ value according to the dispersion relations in the previous subsection are listed in Table~\ref{tab:rates}.
%We see that Im[$-\omega_1$], Im[$-\omega_2$], and Im[$-\omega_3$] are almost the same as $-\gamma_n$, $-\Gamma_{ambi}$, and $-\Gamma_{re}$, respectively, and therefore correspond to the the damping processes due to ion-neutral collisions, ambipolar diffusion, and recombination, respectively. Although the result is expected from the dispersion relations for the post-shock region, a comparison of Table~\ref{tab:eigen} and Table~\ref{tab:rates} suggests that the aforementioned three damping modes also exist inside the shock. 

After identifying the modes and determining their underlying physics through the simple dispersion relations, we study the exact solutions of the entire set of linearized equations in Equation~(\ref{eq:disp_matrix}). 
%\textbf{\textcolor{blue}{The left panel of Figure~\ref{fig:eigenvalue} shows}} the eigenvalues inside the C-shock at $x=0.1$, $x=$\textbf{\textcolor{blue}{0.2}}, and 0.3 pc as well as the eigenvalues in the post-shock region at $x=0.5$ pc. 
Figure~\ref{fig:eigenvalue} shows the properties of the eigenvalues, which describe the behaviors of the growth/damping rates (left panel) and the wave frequencies for the modes of interest (right panel).
The left panel of Figure~\ref{fig:eigenvalue} overplots the five eigenvalues Im[$|\omega|$] at four different locations (colored dots), both inside ($x=0.1$, 0.2, 0.3~pc) and outside ($x=0.5$~pc) the C-shock, with relevant rates in the system. 
Of the five eigenmodes, only one unstable wave mode exists inside the C-shock (red dots). 
In contrast, 
%all the decaying wave modes exist at $x=0.5$ pc in the post-shock region (i.e. \textbf{\textcolor{blue}{no red but all blue dots}}). 
there are only decaying wave modes (blue dots) in the post-shock region (the $x=0.5$~pc location).

To understand the properties of these eigenvalues, the left panel of Figure~\ref{fig:eigenvalue} also shows the rates relevant to the different decaying modes according to the dispersion relations in the previous subsection.
We see that  the first three largest Im[$|\omega|$] are almost the same as $\gamma_n$, $\Gamma_{ambi}$, and $\Gamma_{re}$ and therefore correspond to the the damping processes due to ion-neutral collisions, ambipolar diffusion, and recombination, respectively. Although the result is expected from the dispersion relations for the post-shock region, the left panel of Figure~\ref{fig:eigenvalue} suggests that the aforementioned three damping modes also exist inside the shock. 
In addition, Figure~\ref{fig:eigenvalue} also illustrates that the two smallest Im[$|\omega|$] at $x = 0.1$, 0.2, and 0.3 pc are consistent with Re[$\Gamma_{GLV}$] for the pair of the growing and decaying wave modes.

%Furthermore, Im[$-\omega_4$] and Im[$-\omega_5$] are close to $\gamma_i$ and $\Gamma_{th}^2/\gamma_i$, respectively, at $x=0.5$ pc; thus these eigenvalues are \textbf{\textcolor{blue}{more}} consistent with the dispersion relations for $\gamma_i > \Gamma_{th}$ \textbf{\textcolor{blue}{in the post-shock region as expected from the previous subsection}}. However, values at $x =$ 0.1 and 0.3 pc differ from  $-\gamma_i$ and $-\Gamma_{th}^2/\gamma_i$.
%%whereas they differ at $x=0.1$ and $0.3$ pc.  
%When $k$ is increased such that $\gamma_i \ll \Gamma_{th}$, Im[$-\omega_4$] and Im[$-\omega_5$] become close to $\gamma_i/2$ in the post-shock region (not shown), in accordance with the dispersion relation \textbf{\textcolor{blue}{described by Equation(\ref{eq:post_decay2})}}.
%The aforementioned  results infer that 
%these two decaying modes in the post-shock region replace the overstable mode and its counterpart of decaying mode inside the shock (Equation\ref{eq:disp_Gu}). We trace the evolution of the eigenvalue and eigenmode from the in-shock to the post-shock region around $x\approx 0.4$ pc. We find that when the overstable mode propagates to the post-shock region, it gradually transforms into a decaying mode with the damping rate of $\approx \Gamma_{th}^2/\gamma_i$ in our fiducial case. 
%with the damping rate $\approx \Gamma_{tj}^2/\gamma_i$ if $\Gamma_{th} < \gamma_i$.

Furthermore, the left panel of Figure~\ref{fig:eigenvalue} shows that the last two smallest Im[$|\omega|$] are close to $\gamma_i$ and $\Gamma_{th}^2/\gamma_i$ at $x=0.5$ pc; thus these eigenvalues are consistent with the dispersion relations for $\gamma_i > \Gamma_{th}$ in the post-shock region, as expected from the previous subsection.
%differ from  $-\gamma_i$ and $-\Gamma_{th}^2/\gamma_i$.
%whereas they differ at $x=0.1$ and $0.3$ pc.  
When $k$ is increased such that $\gamma_i\approx \Gamma_{th}$ or even $\gamma_i \ll \Gamma_{th}$, the decaying rate of the modes with the last two smallest Im[$\omega$] 
becomes close to $\gamma_i/2$ in the post-shock region (not shown), in accordance with the dispersion relation  described by Equation~(\ref{eq:post_decay2}) or the more general form $\Gamma=-\gamma_i \pm \sqrt{\gamma_i^2-4\Gamma_{th}^2}/2$ shown in Section \ref{sec:dispersion}.
%\textbf{\textcolor{blue}{On the other hand, from the same figure, the two smallest Im[$|\omega|$]}} at $x =$ 0.1, {\bf 0.2}, and 0.3 pc \textbf{\textcolor{blue}{are consistent with Re[$\Gamma_{GLV}$] for the pair of the growing and decaying wave modes.}} 
Combined with the fact that these two modes represent the pair of the growing and decaying wave modes within the C shock (see Equation~(\ref{eq:disp_Gu})), this suggests that
these two decaying modes in the post-shock region replace the overstable mode and its counterpart of decaying mode inside the shock. We trace the evolution of the eigenvalue and eigenmode from the in-shock to the post-shock region around $x\approx 0.4$ pc. We find that when the overstable mode propagates to the post-shock region, it gradually transforms into a decaying mode with the damping rate of $\approx \Gamma_{th}^2/\gamma_i$ in our fiducial case. 
%with the damping rate $\approx \Gamma_{tj}^2/\gamma_i$ if $\Gamma_{th} < \gamma_i$.

Figure~\ref{fig3} illustrates 
the physical properties of the unstable mode as a function of $x$ in the comoving frame of the neutrals.  The left panel of Figure~\ref{fig3} displays the growth rate $\Gamma_{grow}$ ($=-{\rm Im}[\omega]>0$) 
%and show the slowest decay rate of the mode ($\Gamma_{grow}<0$)  outside the C-shock. 
and wave frequency $\omega_{wave,n}$ ($={\rm Re}[\omega]+kV_n$) of the unstable mode in the comoving frame of the neutrals.  The growth rate Re[$|\Gamma_{GLV}|$] is also plotted in this panel for comparison. According to the simplified dispersion relation in Equation~(\ref{eq:disp_Gu}), Re$[|\Gamma_{GLV}|]= \omega_{wave,n}= \Gamma_{grow}$. These parameters are not completely identical for the exact solutions shown in left panel of Figure~\ref{fig3}, however;
$\omega_{wave,n}$ and Re$[|\Gamma_{GLV}|]$ are marginally larger than $\Gamma_{grow}$ due to the presence of less dominant terms that are not significantly smaller than Re[$|\Gamma_{GLV}|$], such as $\gamma_i$ and $\Gamma_{th}$ (left panel of Figure~\ref{fig2}). Owing to the same reason, the phase difference between $\delta v_n$ and $\delta \rho_n$ of the unstable model is not exactly $(3/4)\pi$, as expected from the dispersion relation, but approaches this toward $\approx 0.85\pi$ from the shock boundaries to the middle of the shock width, as depicted in the right panel of Figure~\ref{fig3}. The same panel also shows that $\delta \rho_n$ and $\delta \rho_i$ are almost in phase due to the ionization-recombination equilibrium. When we remove the ionization and recombination terms in the linearized equations, the unstable mode almost disappears in the eigenvalue problem.
Consequently, the overall results are consistent with the results expected from the simple dispersion relation for the drag instability. The physical picture is that in the comoving frame of the neutrals, the ions drift toward the shock front at $x=0$ pc ($V_d<0$), and the wave travels toward the shock front as well ($\omega_{wave,n}>0$) with a phase of $\delta v_n$, which leads $\delta \rho_n$ by approximately $(3/4)\pi$.

Because the overstable mode propagates inside the shock and subsequently decays in the post-shock region, a problem arises regarding whether sufficient time is available for the unstable wave to grow. The right panel of Figure~\ref{fig:eigenvalue} shows the smallest Im[$|\omega|$] (dots) and its corresponding wave frequency Re[$-\omega$] (crosses) inside the C-shock at $x=0.1$, 0.2, and 0.3 pc 
and in the post-shock region at $x=0.5$ pc for our fiducial model with $k = 1/$(0.015 pc).
The ratio Re[$-\omega$]/Im[$|\omega|$] is approximately 10–20, which indicates that unstable waves travel downstream approximately 10–20 times faster than their growth rate. This high rate of wave propagation in the shock frame is caused by the advection of waves by the fast downstream motion with a speed of $V_n$, which dominates over $v_{ph,n}$ of the unstable mode. The aforementioned statement is verified by the result 
Re$[-\omega]$ $\approx kV_n$,
as presented in the right panel of Figure~\ref{fig:eigenvalue}. In particular, Re$[-\omega]$ is exactly equal to  $kV_n$ in the post-shock region, which agrees with the dispersion relations. 
%This indicates that when $\gamma_i \gg \Gamma_{th}$ in the post-shock region (left panel of Figure 3), no waves but decaying modes exist in the comoving frame of the flow. 
In the following section, we investigate whether the shock width is sufficient or if any favorable pre-shock conditions exist for the fast-traveling wave to grow significantly.

\section{Total GROWTH of an UNSTABLE WAVE over the shock width}
\label{sec:growth}

\subsection{The Maximum-Growing Mode}

In the preceding WKBJ analysis, we simply kept $k$ constant in the fiducial case to study the basic properties of a local unstable/decaying mode.
To examine whether an unstable wave mode can grow appreciably over the shock width,
%until it propagates to the post-shock region where the wave succumbs to damp. 
we consider a mode of a given wave frequency of $\omega_{wave}=$Re[$\omega$] in the shock frame (e.g., $\sim-1$e$-11$~s$^{-1}$ according to %Table~\ref{tab:eigen}).
the right panel of Figure~\ref{fig:eigenvalue}).
Equation(\ref{eq:disp_matrix}) is solved for both the growth rate $\Gamma_{grow}$ ($\equiv$ Im[$\omega]$ when Im[$\omega]<0$ or zero otherwise)  and wavenumber $k$ corresponding to a given $\omega_{wave}$ everywhere in the shock. The perturbation amplitude $|U|$ is arbitrary in a linear analysis for normal modes. For our purpose of evaluating the global growth, we can gain a general sense of the total growth of the unstable model $U$ by setting its norm equal to 1 everywhere in the shock. The local growth of the unstable wave mode is $\exp[\Gamma_{grow} (dx/v_{ph})]$, where $v_{ph}$ is the phase velocity of the wave in the shock frame and is equal to $-\omega_{wave}/ k$. Consequently, the 
total growth of the mode can be computed by integrating the local growth over the entire shock width (i.e., $\exp(\int_{\rm shock\ width} \Gamma_{grow} dx/v_{ph})$). We vary the wave frequency $\omega_{wave}$ and repeat the above procedure to seek the particular mode with the maximum total growth (MTG). MTG$=1$ when no growth occurs (i.e., $\Gamma_{grow}=0$ everywhere).

\begin{figure}
\plottwo{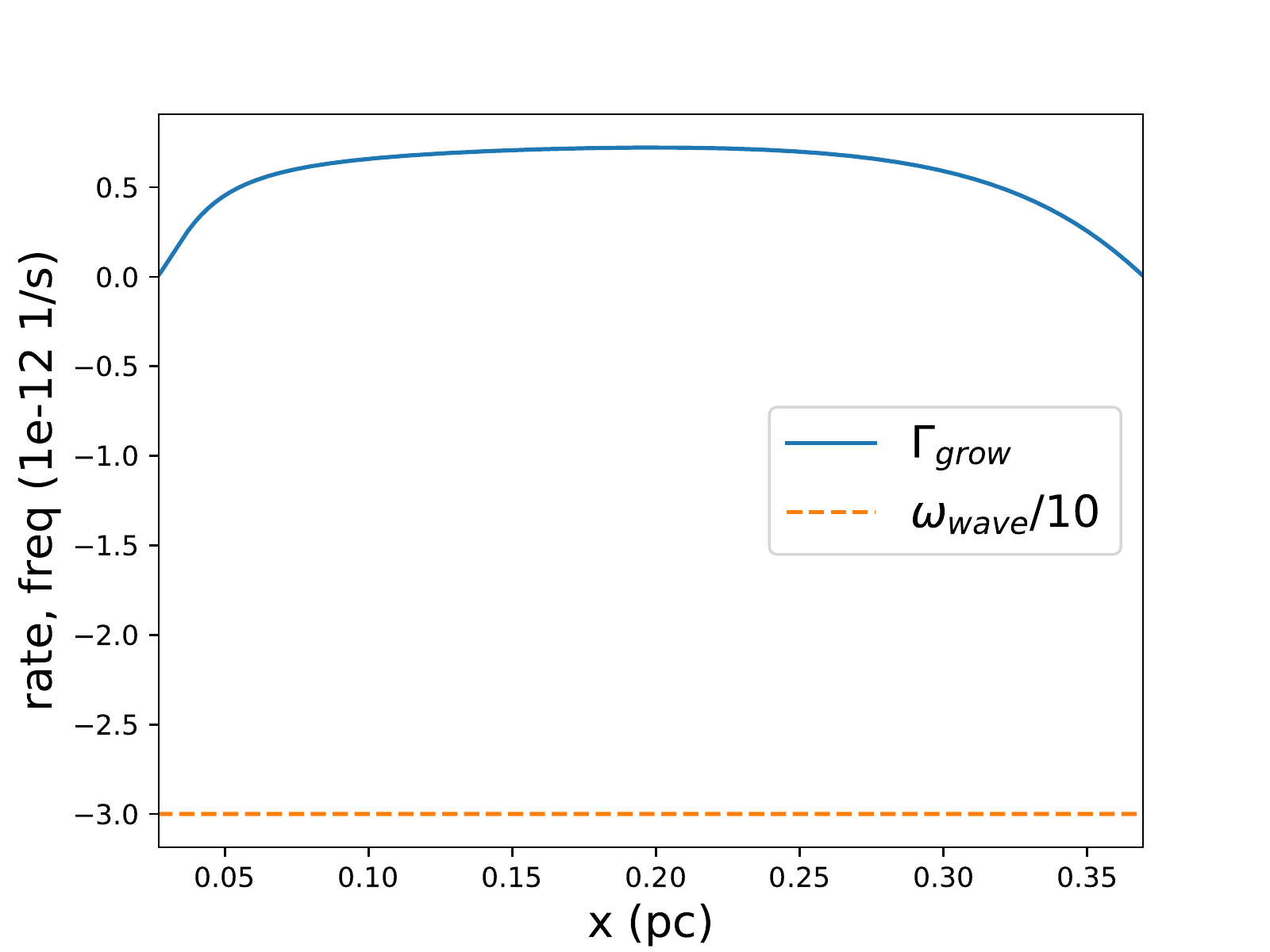}{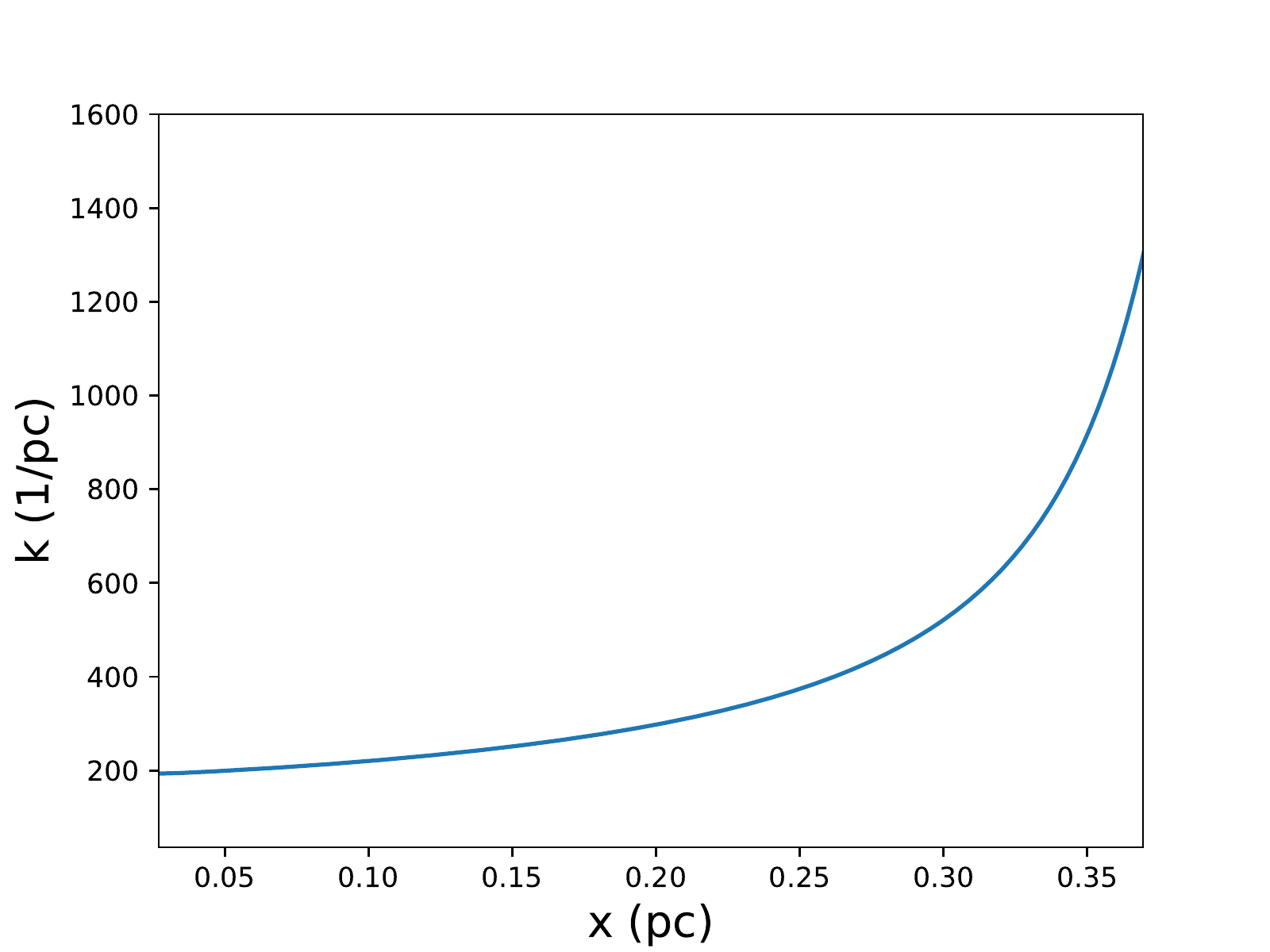}
\caption{Profiles of the growth rate $\Gamma_{grow}$ (left panel) and the corresponding wavenumber $k$ (right panel) of the unstable mode with the wave frequency ($\omega_{wave}$) of -3e$-11$~s$^{-1}$, which is indicated by the dashed orange line in the left panel.}
\label{fig4}
\end{figure}

In our fiducial model for the steady C-shock, the mode $\omega_{wave}\approx-3$e$-11$~s$^{-1}$ is responsible for the MTG. The mode properties as a function of $x$  
are shown in Figure~\ref{fig4}. The figure indicates that the growth rate $\Gamma_{grow}$
increases from the shock front around $x\gtrsim 0.027$ pc, declines after the midpoint of the shock width, and then drops to zero at the end of the shock width at approximately $x= 0.37$ pc. The parameter $\Gamma_{grow}$ is more than 10 times smaller than $\omega_{wave}$. Nevertheless, the wavenumber $k$ 
%remains constant values in the pre- and post-shock regions but 
increases  downstream across the shock width due to the gradient of the background states. The result can be explained by the relation Re$[-\omega] \approx kV_n$  as discussed in the preceding section. Because the wave frequency $\omega_{wave}$ ($=$Re$[\omega] \approx -kV_n$) is approximately constant for a given wave mode, $V_n$ decreases and hence $k$ increases with $x$ as the neutrals are compressed and thus decelerated downstream inside the shock.

\begin{figure}
\plottwo{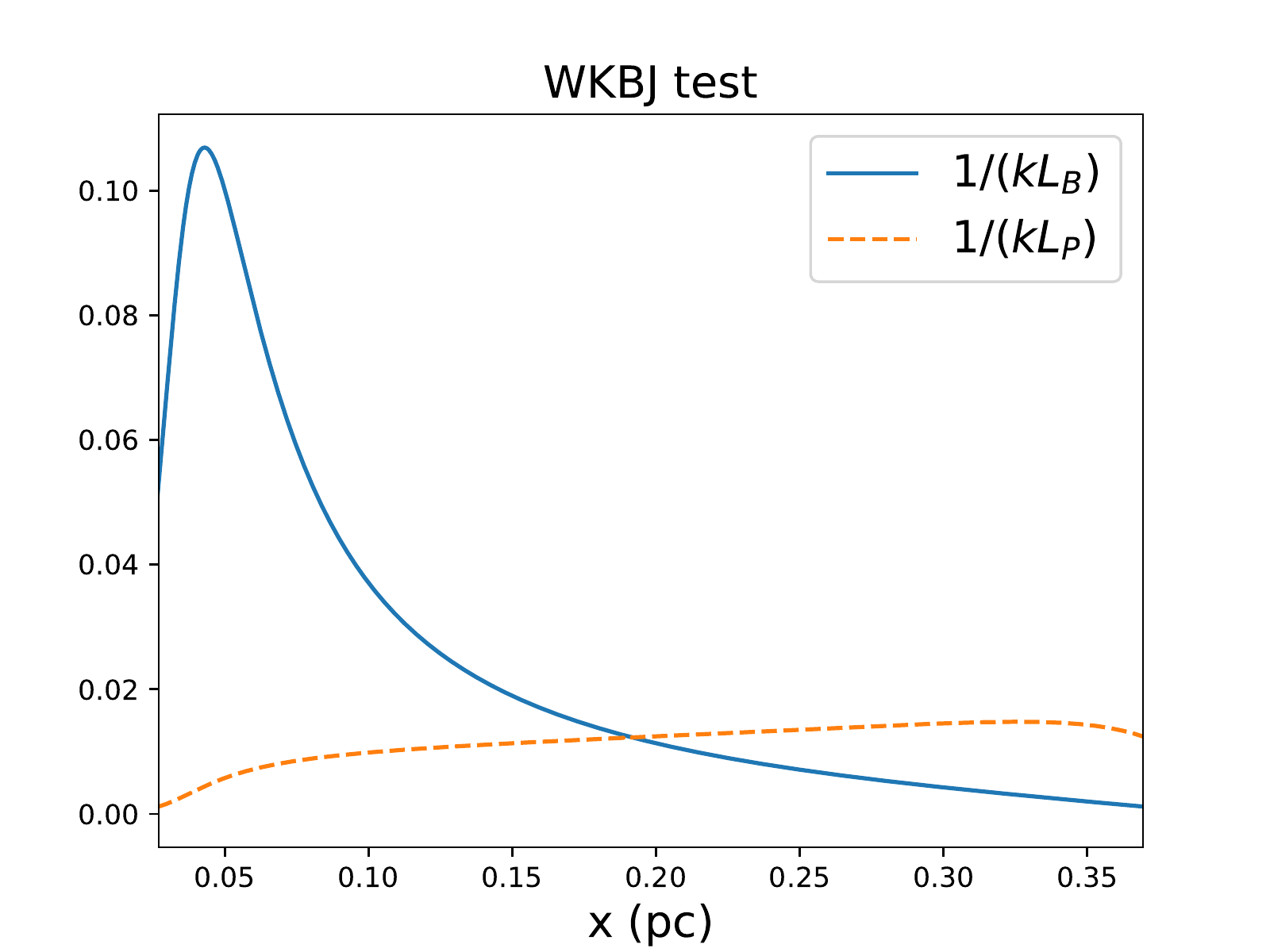}{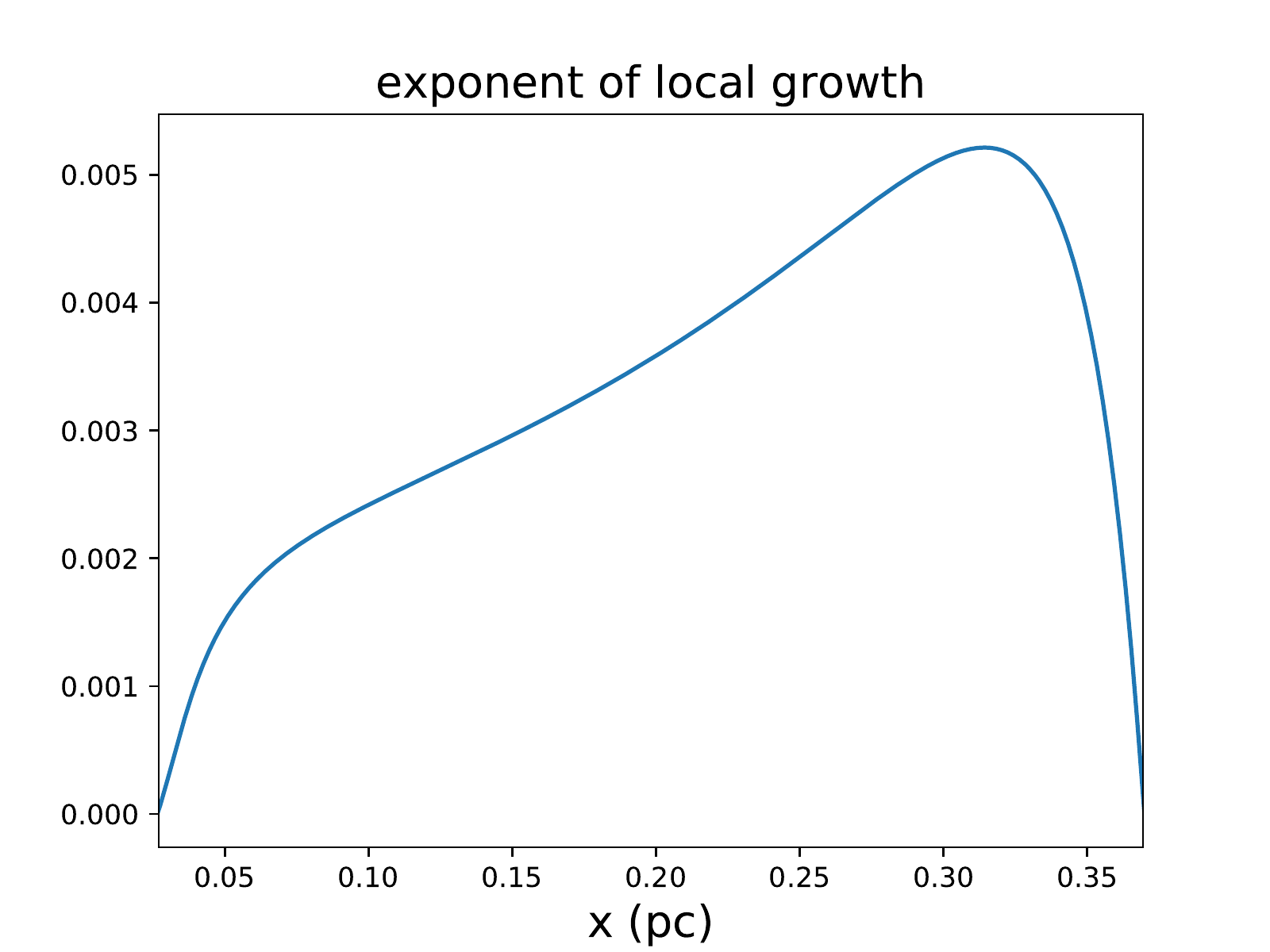}
\caption{WKBJ test for the unstable mode with $\omega_{wave}=-3$e$-11$~s$^{-1}$ (left panel) and the profile of the exponent of the local growth (right panel) in the fiducial model.}
\label{fig5}
\end{figure}

The left panel of Figure~\ref{fig5} indicates that $1/(kL_B)$ and $1/(kL_p)$ of the unstable mode are considerably smaller than 1, which justifies the WKBJ approximation for the calculations. The right panel of Figure~\ref{fig5} displays the profile of the exponential exponent of the local growth ($\Gamma_{grow} dx/v_{ph}$) across the shock. The profile indicates that the unstable mode gains more growth in the rear of the shock transition because the mode has a larger $k$ farther downstream within the shock and hence propagates more slowly to allow for more growth.
This maximum-growing mode caused by the drag instability results in an MTG value of approximately $9.9$ within the steady C-shock, which implies that an initial perturbation of finite magnitude (i.e., $\delta \rho_n/\rho_n \sim 1/10$) is required for a substantial growth of the mode to the nonlinear regime. 

\subsection{A Parameter Study}
\label{sec:parameter}

In addition to the fiducial case, we also calculate the MTG for the C-shock models listed in Table 1 of \citet{CO12},
which represent various conditions for C-shocks to form in star-forming clouds.
%\textbf{\textcolor{blue}{where the authors focused on the typical parameters for C-shock systems within star-forming molecular clouds}}. 
The results of the parameter study are presented in Table~\ref{tab:grow_tot}. In accordance with \citet{CO12}, the letters N, V, B, and X of the model names denote the variations in the $n_0$, $v_0$, $B_0$, and $\chi_{i0}$ values of the models, respectively. In this study, the variation in $\chi_{i0}$ is simply calculated by changing the recombination rate $\beta$ while maintaining the ionization rate $\xi_\mathrm{CR}$ constant. Table~\ref{tab:grow_tot} is almost identical to Table 1 of \citet{CO12}, except for the last two columns, which present the MTG and the corresponding $\omega_{wave}$ of the unstable mode in each model. 

Table~\ref{tab:grow_tot} indicates that for the unstable modes with the MTG, $\omega_{wave}$ has low model dependence and has the value of approximately 1-4e$-11$~s$^{-1}$. In Table~\ref{tab:grow_tot}, the MTG exhibits an increasing trend with the increasing $n_0$ and $v_0$ but decreasing $B_0$ and $\chi_{i0}$. A wider shock width $L_{shock}$ does not necessarily result in a larger MTG. The MTG changes by approximately 10-20 times due to the variations in $n_0$, $B_0$ and $\chi_{i0}$ in the parameter study, resulting in a modest mode growth as that in the fiducial model. By comparison, the MTG is more sensitive to the change in $v_0$. In the parameter study, the stronger shocks characterized by higher shock speeds ($v_0=8$-12 km/s)  with wider shock widths ($L_{shock}\gtrsim 2$ pc), as presented in Models V08, V10, and V12, can boost the MTG to a value of several hundreds.

In addition to considering the original models in Table 1 of \citet{CO12}, we run two additional models, which are denoted as ``Fig4CO12" and ``Fig5CO12" in the last two rows of Table~\ref{tab:grow_tot}. These two models correspond to the scenarios shown in Figures 4 \& 5 of \citet{CO12} for their 1D MHD simulations, which represent the interesting cases of weak C shocks with $v_0=1$ km/s.
%(i.e., $\mathcal{M_A}\approx 6$). 
The MTG values of these two cases are almost 1, meaning little to no growth of the initial perturbation for weak C shocks.

The overall trend of the change in the MTG with the pre-shock parameters $n_0$, $v_0$, $B_0$, and $\chi_{i0}$ may be qualitatively but not quantitatively understood as follows. The exponent of the MTG (i.e., 
$\int_{shock\ width} \Gamma_{grow} dx/v_{ph})$) can be approximated to $ t_{ad} \Gamma_{grow}$ where $t_{ad}$ is the ambipolar drift timescale across the shock width, which is equal to $L_{shock}/V_d$. Inside the shock, the ions are compressed prior to the neutrals. Thus, we consider $r_n \sim 1$ and $r_B \sim r_f$, where $r_f$ is the final compression ratio, which is proportional to $v_0/V_{A,n,0}$ \citep{CO12}. Using the dispersion relation $\Gamma_{grow} \sim \sqrt{k/L_B} V_{A,n} \approx \sqrt{\omega_{wave}/(L_B V_n)} V_{A,n}$ as well as  $L_B\sim L_{shock}$, $V_d \sim V_n \sim v_0/r_f$, and $V_{A,n} \propto B n_n^{-1/2} \propto (r_B B_0)(r_n n_0)^{-1/2}$, we obtain the following relations: $V_{A,n} \sim v_0$ and $V_d \sim V_{A,n,0}$. Because $L_{shock} \propto n_0^{-3/8} v_0^{1/4} B_0^{1/4}  \chi_{i0}^{-1/2}$ \citep{CO12}, it follows that $t_{ad} \Gamma_{grow} \propto L_{shock}^{1/2} V_d^{-3/2} v_0 \propto n_0^{3/8} v_0^{5/4} B_0^{-5/4} \chi_{i0}^{-1/2}$. The scaling result for the exponent qualitatively gives the trend that MTG increases with $n_0$ and $v_0$ but decreases with $B_0$ and $\chi_{i0}$. In terms of a rough physical picture, the ion-neutral drift velocity, which is comparable to the neutral Alfv\'en velocity in the pre-shock region,  is enhanced by strong magnetic fields or a low neutral density, leading to a short ambipolar drift time $t_{ad}$ for the unstable wave to grow. Hence, a positive correlation exists between the MTG and the neutral density $n_0$, whereas a negative correlation exists between the MTG and the magnetic fields $B_0$. In addition, the neutral Alfv\'en speed in the shock is enhanced in a strong shock to result in a high growth rate; that is, a positive correlation between the MTG and the shock speed $v_0$. Finally, $\chi_{i0}$ changes the MTG by varying the shock width. A large $\chi_{i0}$ leads to a wide shock width, which allows more time for the unstable mode to grow.
We discuss the astronomical implication of these results in Section~\ref{sec:grav}.

%2e-11 25
%3e-11 37.8

%We also calculate $\Gamma_{tot,max}$ for the two C-shock models presented in Figure 2 ($n_0=500$ cm$^{-3}$, $v_0=5$ km/s, $B_0=5\mu$G) and Figure 3 ($n_0=500$ cm$^{-3}$, $v_0=5$ km/s, $B_0=10\mu$G) of \citet{CO12}.  In the first model, $\Gamma_{tot,max}\approx 10.6$ for the mode of $\omega_{wave}=-2.3$e$-11$ 1/s. In the second model, $\Gamma_{tot,max}\approx 10.2$ for the mode of $\omega_{wave}=-2.5$e$-11$ 1/s. 

\begin{deluxetable*}{cccccccc}
%	\tablenum{2}
	\tablecaption{MTG and corresponding mode frequency $\omega_{wave}$ of the drag instability in the steady C-shock models of \citet{CO12}. The final two rows correspond to the two additional models used for Figures 4 \& 5 in \citet{CO12}, respectively. The parameter $L_{shock}$ is the shock width estimated using Equation~(42) in \citet{CO12}. \label{tab:grow_tot}}
	\tablewidth{0pt}
	\tablehead{
		\colhead{model} & \colhead{$n_0$ (cm$^{-3}$)} & \colhead{$v_0$ (km/s)}  &
		\colhead{$B_0$ ($\mu$G)} & \colhead{$\chi_{i0}$} &   \colhead{$L_{shock}$ (pc)} & \colhead{$\omega_{wave}$ (s$^{-1}$)} &\colhead{MTG}
	}
	\startdata
	N01 & 100 & 5 & 10 & 5 & 3.15 & $-2$e$-11$ & 10.9  \\
	N03 & 300 & 5 & 10 & 5 & 1.38 & $-3$e$-11$ & 19.1 \\
	N05 & 500 & 5 & 10 & 5 & 0.94 & $-3$e$-11$ & 22.0 \\
	N08 & 800 & 5 & 10 & 5 & 0.66 & $-4$e$-11$ & 23.8 \\
	N10 & 1000 & 5 & 10 & 5 & 0.56 & $-4$e$-11$ & 24.7 \\
	V04 & 200 & 4 & 10 & 5 & 1.68 & $-3$e$-11$ & 6.4 \\
	V06 & 200 & 6 & 10 & 5 & 2.05 & $-2$e$-11$ & 36.0 \\
	V08 & 200 & 8 & 10 & 5 & 2.37 & $-2$e$-11$ & 121.2 \\
	V10 & 200 & 10 & 10 & 5 & 2.65 & $-4$e$-11$ & 327.3 \\
	V12 & 200 & 12 & 10 & 5 & 2.90 & $-3$e$-11$ & 816.2 \\
	B02 & 200 & 5 & 2 & 5 & 0.84 & $-2$e$-11$ &  27.7 \\
	B04 & 200 &  5  &  4  & 5  & 1.18 & $-2$e$-11$ & 25.3 \\
	B06 & 200 &  5  &  6  & 5  & 1.45 & $-2$e$-11$ & 22.5 \\
	B08 & 200 &  5  &  8  & 5  & 1.68  & $-2$e$-11$ & 19.5 \\
	B10 & 200 &  5  &  10  & 5  & 1.87  &  $-2$e$-11$ &  16.6 \\
	B12 & 200 &  5  &  12  & 5  & 2.05  &  $-2$e$-11$ &  13.7 \\
	B14 & 200 &  5  &  14  & 5  & 2.22  &  $-2$e$-11$ &  11.2 \\
	X01 & 200 &  5  &  10  & 1  & 9.37  &  $-2$e$-11$ &  23.5 \\
	X06 & 200 &  5  &  10  & 6  & 1.56  &  $-2$e$-11$ &  14.7 \\
	X10 & 200 &  5  &  10  & 10  & 0.94  &  $-2$e$-11$ &  8.9 \\
	X15 & 200 &  5  &  10  & 15  & 0.62  &  $-2$e$-11$ &  5.1 \\
	X20 & 200 &  5  &  10  & 20  & 0.47  &  $-1$e$-11$ &  3.6 \\
	Fig4CO12 & 200 & 1  &   2  & 10 & 0.18 &   $-8$e$-12$ &  1.05 \\
	Fig5CO12 & 500 &  1  &  4  & 10  &  0.13  & N/A  &  1 \\
%	0.6 & $2$e$-06$ & $6.135$e$-11$ & $2.446$e$-10$ & $2.133$e$-12$ & $1.97$e$-13$ \\ 
	\enddata
	
%	\tablecomments{This table shows the notations and parameters used in this article; notice that some values change with the stellar mass}
\end{deluxetable*}

\section{Discussion}
\label{sec:disc}

In this work, we study the drag instability in non-self-gravitating, steady-state 1D C-shocks under particular conditions representative of the condition in turbulent star-forming molecular clouds. In this section, we discuss the possible behaviors of the drag instability in numerically evolving C-shocks (Section~\ref{sec:sim}) and in the physical space (i.e.,~under the existence of self-gravity; Section~\ref{sec:grav}) to further explore the practical applications of the drag instability.
Because there is currently no direct observational evidence of C-shocks in turbulent clouds (see the references in Section~\ref{sec:intro}), we frame our arguments using numerical C-shocks and/or the typically observed properties of the parent clouds wherein the C-shocks form as the general guidance. 

We also note that 
it is currently not clear whether the drag instability can occur in oblique C-shocks \citep[see, e.g.,][]{Wardle1991,MacLow95,Ashmore10,CO12}, which requires one more dimension than our 1D analysis here.
%We recall the streaming instability mentioned in Section~\ref{sec:intro} as an example of fluid instability driven by drag force. 
The drag instability can occur in 1D systems because of the ionization and recombination terms in the continuity equation of ions (see Equation~(\ref{eq:2})). These source terms facilitate the growth of density clumps in 1D via drag in the absence of magneto-acoustic modes and another dimension. In contrast, it has been known that the ionization equilibrium precludes the Wardle instability (a 2D/3D effect) in C-shocks \citep{Wardle,MacLow97,Stone}.
Analogously, it is worth noting that as an incompressible mode, the streaming instability (see Section~\ref{sec:intro}) is prohibited in a 1D flow \citep{YG05}. 
We thus restrict our discussions below to 1D systems alone.

\subsection{Drag Instability in C-shock Simulations}
\label{sec:sim}

While the model of the drag instability was developed based on the steady-state profile of C-shocks, it is possible that the drag instability could occur in time-dependent simulations of C-shocks. Conceptually, this could occur when the evolving C-shock system is very close to, but not exactly equal to, the steady-state C-shock structure. If the deviation from the steady-state profile happens to satisfy the unstable mode favored by the drag instability, this local perturbation could evolve and grow with time.

In addition to deriving the structure of a steady-state C-shock, \citet{CO12} also numerically obtained the C-shock structure by simulating two colliding flows using the \textsc{Athena} code \citep{Stone08}. Instead of computing two fluids comprising the ions and neutrals, as shown in Equations~(\ref{eq:1})--(\ref{eq:5}), \citet{CO12} simulated the equations for the neutrals alone under the strong-coupling approximation, that is, ${\bf f_d} = {\bf f_L} =(1/4\pi) (\nabla \times {\bf B})\times {\bf B}$. Therefore the two-fluid equations considered in this study can be reduced to the following one-fluid equations for the neutrals \citep[see also e.g.,][]{MacLow95}:
\begin{eqnarray}
{\partial \rho_n \over \partial t}+ \nabla \cdot (\rho_n {\bf v_n})=0 \label{eq:n_mom}\\
\rho_n \left[ {\partial {\bf v_n} \over \partial t} + ({\bf v_n} \cdot \nabla ) {\bf v_n} \right]+\nabla p_n = {1\over 4\pi}(\nabla \times {\bf B})\times {\bf B}\label{eq:n_mass}\\
{\partial {\bf B} \over \partial t} + \nabla \times ({\bf B} \times {\bf v_n})= \nabla \times \left\{ {\bf B} \times  \left[ {1\over4 \pi  \gamma \rho_i \rho_n}   (\nabla \times {\bf B})\times {\bf B} \right] \right\}.\label{eq:ind_AD}
\end{eqnarray}
Note that the momentum equation (Equations~(\ref{eq:n_mom})) and the mass conservation equation (Equations~(\ref{eq:n_mass})) for neutrals are identical to that in the ideal MHD limit (see Equations~(\ref{eq:1}) and (\ref{eq:3})), but the induction equation (Equations~(\ref{eq:5})) now has a correction term from the ion-neutral drift (see also Equation~(46) in \citealt{CO12}). 
Although 
the numerical results of \citealt{CO12} were consistent with the analytical expectation for the C-shock structure, no instabilities were observed in their simulations. This result contrasts with the result of our linear analysis. 
We discuss possible explanations below.

Because the drag instability is derived from the two-fluid model in this study,
we examine whether the strong-coupling approximation can dismiss the drag instability in the one-fluid model adopted in \cite{MacLow95} and \cite{CO12}, for example. In the two-fluid model, the drag instability arises from the perturbed drag term $\gamma \rho_i V_d \delta \rho_i/\rho_i$ in the momentum equation for the neutrals (see Equation~(\ref{eq:n_drag})). In the one-fluid model, this term is replaced by the perturbation of magnetic pressure $-(V_{A,n}^2/B) d \delta B/dx=-ik(V_{A,n}^2/B) \delta B$, which is in turn linked to the perturbation of the diffusion-corrected induction equation (Equation~(\ref{eq:ind_AD})):
\begin{equation}
\Gamma \delta B +\rmi k \delta v_n B \approx {V_{A,n}^2 \over \gamma \rho_i} \left( {d^2 \delta B \over dx^2} -{1\over \rho_n} {d \delta \rho_n \over dx} {dB \over dx} \right)
=-k^2{V_{A,n}^2 \over \gamma \rho_i} \delta B+\rmi k V_d {\delta \rho_n \over \rho_n} B. \label{eq:ind}
\end{equation}
Note that we have adopted the WKBJ approximation but still retained the term with $dB/dx$ due to a large $V_d$ ($\propto dB/dx$) for our interest.  The two terms from the right-hand side of Equation~(\ref{eq:ind}) arise from the perturbation of the ambipolar diffusion. The first term is the typical diffusion term $k^2 D_{ambi}$ with the ambipolar diffusivity $D_{ambi} \equiv V_{A,n}^2/(\gamma \rho_i)$.
Rearranging the above equation, we find
\begin{equation}
-\rmi k V_{A,n}^2 {\delta B \over B} \approx \gamma \rho_i V_d {\delta \rho_n \over \rho_n} - \gamma \rho_i \delta v_n + \rmi {1\over k} \Gamma \gamma \rho_i {\delta B\over B},
\end{equation}
where the first term on the right-hand side (i.e. the last term on the right-hand side of Equation~(\ref{eq:ind})) is the term required for the drag instability in the two-fluid model.
This suggests that the drag instability can occur in the strong-coupling limit.

Indeed, the ion-neutral drift in some of the simulated C-shocks may not be 
%The other possibilities would be that the C shock is not
sufficiently strong to either initiate the drag instability (models Fig4CO12 and Fig5CO12 in  Table~\ref{tab:grow_tot}) or generate appreciable growth without an initial perturbation from the background structure (i.e.,~the steady-state solution).
Another possible factor of the missing instability is numerical resolution.
The spacial resolution adopted in \cite{CO12}'s 1D simulations is 0.01 pc, which could be too coarse to resolve the drag instability. 
In fact, as shown in Figure~\ref{fig4}, the wavenumber $k$ corresponds to the unstable mode in our fiducial model is $\gtrsim 500$~pc$^{-1}$ within the C-shock, which suggests that the physical scale of the growing perturbation could be smaller than $0.002$~pc. 
%(refer to the large wavenumber $k$ in Figure ~\ref{fig4}). 
%Also, self-gravity in the overdense region of the growing wave may become potentially important if the perturbations of the drag instability can grow to the nonlinear regime. 

We further note that the drag instability has not been reported in most of the previous studies investigating the 1D C-shock structure \citep[e.g.,][]{SmithML97,Chieze98,Ciolek02,vanLoo09}.
\cite{SmithML97} adopted the so-called frozen-in condition \citep[e.g.,][]{Wardle} assuming ion conservation. This means that ionizations and recombinations are neglected, and the ion number density through the C-shock is purely determined by the compression of the magnetic field via the ion conservation equation and the induction equation. Because the dependence of the ion density on the neutral density is essential for the drag instability to occur, it is not surprising that the drag instability was suppressed in their simulations.
For works that included microphysics and/or the chemistry of the C-shock system \citep[e.g.,][]{Chieze98,Ciolek02,vanLoo09}, the ionization and recombination processes became more complicated and could affect the timescale on which the drag instability grew. We also note that the increased gas temperature from shock compression (up to $\sim 10^2-10^3$~K) may completely prevent the drag instability (which is derived using isothermal equation of state) in these simulations. 

%Also, \cite{SmithML97} noted that, in their 1D, multi-fluid simulations of C shock formation, the accuracy of reproducing the standard C shock profiles highly depends on the grid resolution (see their Figure~2). This could be related to the existence of unstable wave mode from the drag instability.
\subsection{The Significance of the Drag Instability in Astronomical Systems}
\label{sec:grav}

In this work, we study the drag instability in C-shocks with conditions that can arise from clump-clump collisions or cloud-scale supersonic turbulent flows in typical star-forming regions.
One of the most tantalizing questions for shocks in this context is whether a shock instability can lead to fragmentation that is subject to gravitational collapse and eventually induce star formation. 
However, our analysis here focuses on the behavior of the drag instability within the steady-state profiles of C-shocks, which is linear and non-self-gravitating.
%does not include self-gravity.
These linear analyses are therefore not applicable to directly address this issue and predict any nonlinear outcomes.

Nonetheless, the linear theory could still provide indications on the possible consequence of the instability.
The values of the MTG listed in Table~\ref{tab:grow_tot} indicate that under preferred circumstances, a small perturbation from the steady-state solution could lead to large ($\gtrsim 100\times$) growth.
Based on the parameter study, the most favorable condition for drag instability to grow significantly is in strong shocks (high inflow velocities; $v_0 \gtrsim 5$~km/s). This condition is consistent with the typical environment in giant molecular clouds or molecular cloud complexes \citep[velocity dispersion $\sigma_v \sim 1-10$~km/s, see recent observations in, e.g.,][]{Heyer09,Miura12,Evans14,Garcia14,Nguyen16}. 
The total growth induced by the drag instability can also be further enhanced by efficient ambipolar diffusion, i.e.,~weak-coupling between neutrals and ions, and/or relatively low ionization fractions ($n_i/n_n \lesssim 10^{-7}$). 
%which may be typical inside dense cloud cores when high-energy cosmic rays and radiation are blocked due to large extinctions \citep{McKee10}, though it is not clear whether the ionization-recombination equilibrium still holds  
Because the efficiency of nonideal MHD diffusivity is highly dependent on chemical composition and microscopic physical processes, this condition could be typical in some molecular clouds permitted by dust grain properties, e.g.,~regions with larger grains \citep{Nishi91,Nakano02}.
Still, we note that these conditions potentially favored by the drag instability to develop gravitationally unstable structures were derived following the guidance from the 1D linear analysis, and thus may not be applicable to more complex systems.

\begin{figure}
    \plotone{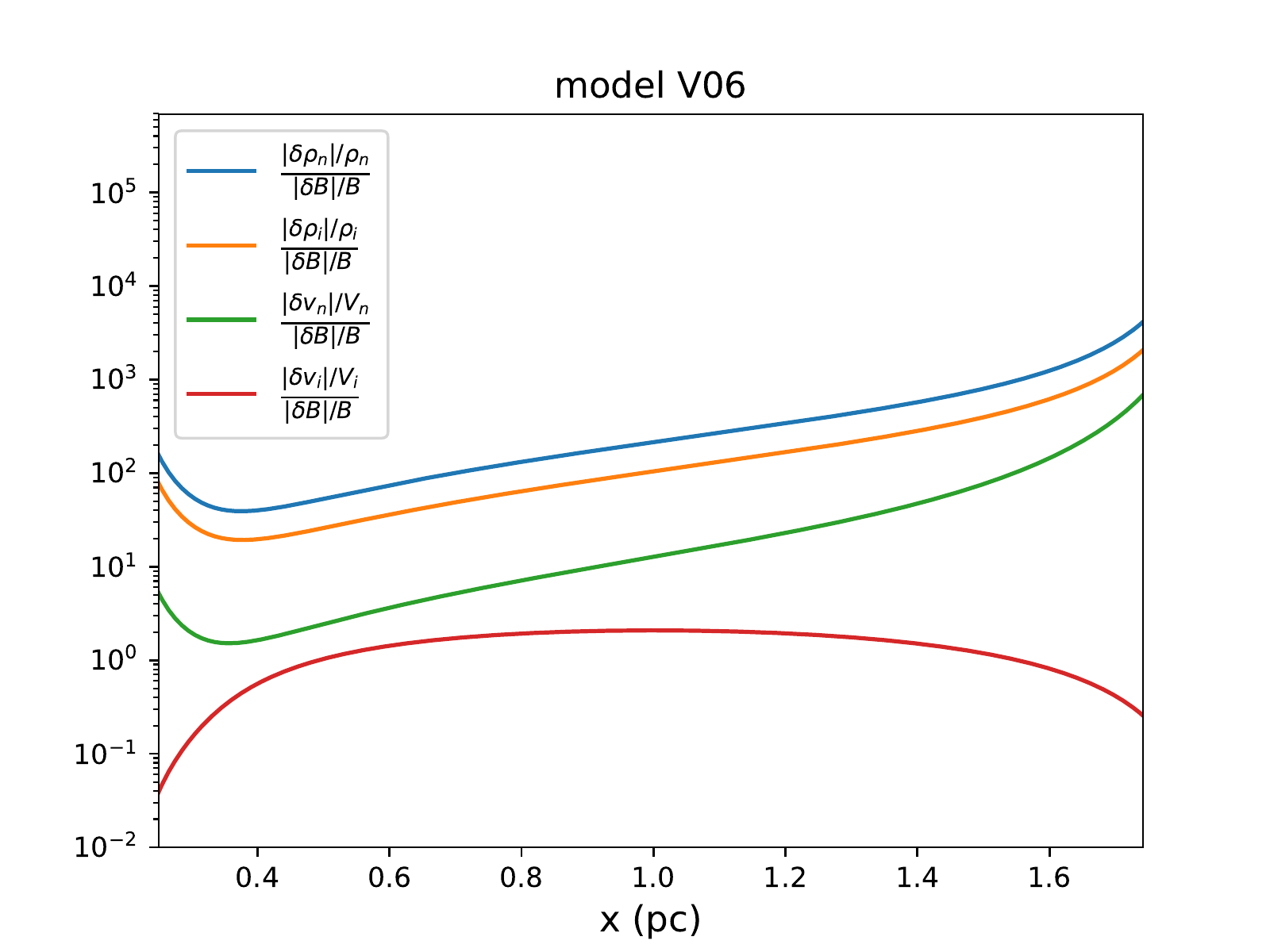}
    \caption{Relative amplitudes of density and velocity perturbations of neutrals and ions (normalized by the magnetic field perturbation) along the C-shock for the growing mode in model V06 (see Table~\ref{tab:grow_tot} for the model parameters). The density perturbations of both neutrals and ions are much larger than the perturbation in velocities and magnetic field, indicating that the density enhancement induced by the drag instability is dynamically important. }
    \label{fig:perb_amp}
\end{figure}

In addition, we investigate the perturbation amplitude induced by the drag instability obtained from our eigenvalue problem (see also, e.g.,~Equation~(\ref{eq:n_cont})), 
which is illustrated in Figure~\ref{fig:perb_amp}.
Because the amplitude ratio between any two of the perturbations (density, velocity, or magnetic field) remains the same in the linear regime as the unstable mode grows, 
we plot the relative amplitude of the perturbations in $\rho_n$, $v_n$, $\rho_i$, and $v_i$ normalized by the perturbation in $B$. 
Figure~\ref{fig:perb_amp} shows these perturbations for the growing mode within the C-shock in model V06, which has a moderate MTG in our parameter study (see Table~\ref{tab:grow_tot}). %The relative amplitudes remain the same as the unstable mode grows in the linear regime. 
It is evident from the figure that the density perturbations ($\delta\rho_n/\rho_n$, $\delta \rho_i/\rho_i$) are much larger than both the velocity and magnetic field perturbations ($\delta v_n/V_n$, $\delta v_i/V_i$, $\delta B/B$) everywhere in the shock. 
%Although we show the result only for model V06, 
%which gives a larger-than-average MTG in our parameter study, 
%the similar result also occurs in the fiducial model. 
Similar results are found in other models, which implies
%It implies 
that as the perturbations grow due to the drag instability, the density perturbation would reach the nonlinear phase faster than other perturbations.
As a result,
the density perturbation driven by the drag instability is dynamically significant, and
these unstable density enhancements induced by the growing wave mode within C-shocks could become gravitationally important in turbulent molecular clouds. Further examinations in numerical simulations will help clarify in which scenario the drag instability would become effective during the star-forming process. 

%\subsection{\bf Relevance to Other Work}

%It is currently not clear whether the drag instability can occur in oblique C shocks \citep[see e.g.,][]{Wardle1991,MacLow95,Ashmore10,CO12}, which requires one additional {\bf dimension} than our 1D analysis here.
%We recall the streaming instability mentioned in Section~\ref{sec:intro} as an example of fluid instability driven by drag force. 
%The drag instability can happen in 1D systems because of the ionization and recombination terms in the continuity equation of ions (see Equation~(\ref{eq:2})). These source terms facilitate the growth of density clumps in 1D via drag in the absence of magneto-acoustic modes and another dimension. In contrast, it has been known that the ionization equilibrium {\bf precludes} the Wardle instability (a 2D{\bf /3D} effect) in C shocks \citep{Wardle,MacLow97,Stone}.
%Analogously, it is worth noting that as an incompressible mode, the streaming instability (see Section~\ref{sec:intro}) is prohibited in a 1D flow \citep{YG05}. 

%\section{Numerical simulation}
\section{Summary}
\label{sec:sum}

Based on the background state of steady C-shocks derived by \citet{CO12}, we conduct a WKBJ analysis and confirm the postulation of \citet{Gu04} that the drag instability in the ISM can occur in a 1D isothermal C-shock where the ion-neutral drift motion is sufficiently high as a result of the compressed magnetic fields within the smooth shock transition.
We first focus on a fiducial case for a C-shock model to study the dispersion relation for the drag instability inside the C-shock. The dispersion relations in the post-shock region are also investigated, which reveal all decaying modes associated with ion-neutral collisions, recombination, ambioplar diffusion, and thermal effect. We then
solve the linear equations  for the exact eigenfrequencies and wavenumber of eigenmodes and identify  their physics based on the dispersion relations for growing and decaying modes throughout the C-shock. In our fiducial case, we find that the growing wave driven by the drag instability propagates downstream and subsequently decays by the slow thermal effect associated with neutral-ion collisions in the post-shock region. 

Because the unstable wave has a considerably higher propagation rate than the local growth rate, the mode with the MTG can be identified as it travels across the entire shock width before it is damped in the post-shock region. In addition to the analysis performed with the fiducial model, we also conduct a parameter study to compute the MTG in numerous C-shock models  corresponding to the turbulent environment in typical star-forming regions with various
pre-shock parameters $n_0$, $v_0$, $B_0$ and $\chi_{i0}$. 
We find that the MTG increases with increasing $n_0$ and $v_0$ but decreases with increasing $B_0$ and $\chi_{i0}$. 
In most cases, the MTG  is typically around 10-30 times larger than the initial perturbation for a modest shock, thus requiring the finite amplitude of the initial perturbation to grow to a nonlinear regime. The drag instability hardly occurs in a weak shock with a shock speed $v_0=1$ km/s. Nevertheless, the MTG can become as large as a few hundred for a strong shock with a $v_0$ value of $\gtrsim 8$ km/s in our parameter studies.

We leave the numerical investigation of the topics discussed in the previous section for future work. 
%We can set up the same numerical simulation as \citet{CO12} but we need to follow the shock motion and study. Moreover, there would be many growing modes in the simulation, thus difficult to compare the result with the linear theory. Question: is there any simple way to verify a predicted unstable wave mode in a numerical simulation? Is it possible to set up a one-fluid simulation with the equilibrium state described by \citet{CO12} and initiate an unstable mode with a particular $\omega_{wave}$ from the linear theory to measure the corresponding growth rate and thus to verify the drag instability? We probably leave this to a future work.
We also have restricted our analysis of the drag instability to a 1D C-shock with a transverse magnetic field for simplicity, which means that the magnetic fields cannot be bent. It will be interesting to extend the work to 2D to study the effect of magnetic tension on the drag instability. A 2D study, both analytically and numerically, will also allow for the investigation of the drag instability in oblique C-shocks following the discussions in \cite{CO14}.
In any case, the analytic solutions of the growth rate and wavenumber in our eigenvalue problem presented in this paper have provided useful information for probing and characterizing the drag instability in simulated time-dependent C-shocks.

%Mathematically, it is difficult to realize how this link can lead back to the drag term $\gamma \rho_i V_d \delta \rho_i/\rho_i$ in the momentum equation for neutrals. Physically, it is the fast drag and ionization/recombination rates against the induction rate to produce the drag instability \citep{Gu04}.

%\url{http://journals.aas.org/authors/aastex.html}.

%\begin{figure}
%\plottwo{TP_ga06.eps}{TP_ga8.eps}
%\caption{T-p profile of the planet down to $p \sim 100$ bars for the two opacity cases: $\kappa_v,\kappa_{th}=$0.004,0.006 cm$^2$/g (left panel, $\gamma \approx 0.67$) and 0.03, 0.0037 cm$^2$/g (right panel, $\gamma \approx 8$). The profile of the atmosphere is shown in blue and that of the interior is plotted in red. }
%\label{fig1}
%\end{figure}

%\begin{figure}
%\includegraphics[width=.5\linewidth]{TP_ga06.pdf}\includegraphics[width=.5\linewidth]{TP_ga8.pdf}
%\caption{T-P profile of the atmosphere for two cases: $\kappa_v,\kappa_{th}=$0.004,0.006 (left panel, $\gamma \approx 0.67$) and 0.03, 0.0037 (right panel, $\gamma \approx 8$).}
%\label{fig1}
%\end{figure}

\acknowledgments
We thank the anonymous referee for a helpful and constructive report, and the colloquium committee at the Institute of Astronomy and Astrophysics in Academia Sinica (ASIAA) for offering the opportunity of initiating this work.
Some of the
numerical work was conducted on the high-performance
computing facility at ASIAA.
P.-G.G. would like to thank Min-Kai Lin and Chien-Chang Yen for informative discussions. 
%Some of the numerical works are conducted on the PC clusters at the Theoretical Institute for Advanced Research in Astrophysics (TIARA). 
P.-G.G. acknowledges support from MOST in Taiwan through the grant MOST 105-2119-M-001-043-MY3.
C.-Y.C. acknowledges support from NSF grant AST-1815784. 


\begin{thebibliography}{}
\bibitem[Ashmore et al.(2010)]{Ashmore10} Ashmore, I., van Loo, S., Caselli, P., et al.\ 2010, \aap, 511, A41
\bibitem[Bai \& Stone(2011)]{BaiStone11} Bai, X.-N., \& Stone, J.~M.\ 2011, \apj, 736, 144
\bibitem[Ballesteros-Paredes et al.(2007)]{BP07} Ballesteros-Paredes, J., Klessen, R.~S., Mac Low, M.-M., et al.\ 2007, Protostars and Planets V, 63
\bibitem[CO12()]{CO12} Chen, C.-Y, \& Ostriker, E. 2012, \apj, 744, 124
\bibitem[Chen \& Ostriker(2014)]{CO14} Chen, C.-Y, \& Ostriker, E. 2014, \apj, 785, 69
\bibitem[Chieze et al.(1998)]{Chieze98} Chieze, J.-P., Pineau des Forets, G., \& Flower, D.~R.\ 1998, \mnras, 295, 672
\bibitem[Ciolek \& Roberge(2002)]{Ciolek02} Ciolek, G.~E., \& Roberge, W.~G.\ 2002, \apj, 567, 947
\bibitem[Crutcher(2012)]{Crutcher12} Crutcher, R.~M.\ 2012, \araa, 50, 29
\bibitem[Dalgarno(2006)]{Dalgarno06} Dalgarno, A.\ 2006, Proceedings of the National Academy of Science, 103, 12269
\bibitem[Dapp et al.(2012)]{Dapp2012} Dapp, W.~B., Basu, S., \& Kunz, M.~W.\ 2012, \aap, 541, A35
\bibitem[Draine(1980)]{Draine80} Draine, B. T. 1980, \apj, 241, 1021
\bibitem[Draine \& McKee(1993)]{DM93} Draine, B. T., \& McKee, C. F. 1993, ARA\&A, 31, 373
\bibitem[Draine et al(1983)]{Draine} Draine, B. T., Roberge, W. G., \& Dalgarno, A. 1983, \apj, 264, 485
\bibitem[Evans et al.(2014)]{Evans14} Evans, N.~J., Heiderman, A., \& Vutisalchavakul, N.\ 2014, \apj, 782, 114
\bibitem[Federrath et al.(2011)]{Federrath11} Federrath, C., Sur, S., Schleicher, D.~R.~G., et al.\ 2011, \apj, 731, 62
\bibitem[Fiedler \& Mouschovias(1992)]{Fiedler92} Fiedler, R.~A., \& Mouschovias, T.~C.\ 1992, \apj, 391, 199
\bibitem[Fiedler \& Mouschovias(1993)]{Fiedler93} Fiedler, R.~A., \& Mouschovias, T.~C.\ 1993, \apj, 415, 680
\bibitem[Flower \& Pineau Des For{\^e}ts(1998)]{Flower98} Flower, D.~R., \& Pineau Des For{\^e}ts, G.\ 1998, \mnras, 297, 1182
\bibitem[Flower \& Pineau Des For{\^e}ts(2010)]{Flower10} Flower, D.~R., \& Pineau Des For{\^e}ts, G.\ 2010, \mnras, 406, 1745
\bibitem[Fukui \& Kawamura(2010)]{Fukui10} Fukui, Y., \& Kawamura, A.\ 2010, \araa, 48, 547
\bibitem[Garc{\'\i}a et al.(2014)]{Garcia14} Garc{\'\i}a, P., Bronfman, L., Nyman, L.-{\r{A}}., et al.\ 2014, \apjs, 212, 2
\bibitem[Gressel et al.(2015)]{Gressel15} Gressel, O., Turner, N.~J., Nelson, R.~P., et al.\ 2015, \apj, 801, 84
\bibitem[GLV())]{Gu04} Gu, P.-G., Lin, D. N. C., \& Vishniac, E. T., 2004, \apss, 292, 261
\bibitem[Guillet et al.(2011)]{Guillet11} Guillet, V., Pineau Des For{\^e}ts, G., \& Jones, A.~P.\ 2011, \aap, 527, A123
\bibitem[Gusdorf et al.(2008)]{Gusdorf08} Gusdorf, A., Cabrit, S., Flower, D.~R., et al.\ 2008, \aap, 482, 809
\bibitem[Hennebelle et al.(2016)]{Hennebelle16} Hennebelle, P., Commer{\c{c}}on, B., Chabrier, G., et al.\ 2016, \apjl, 830, L8
\bibitem[Heyer et al.(2009)]{Heyer09} Heyer, M., Krawczyk, C., Duval, J., et al.\ 2009, \apj, 699, 1092
\bibitem[Hezareh et al.(2010)]{Hezareh10} Hezareh, T., Houde, M., McCoey, C., et al.\ 2010, \apj, 720, 603
\bibitem[Hezareh et al.(2014)]{Hezareh14} Hezareh, T., Csengeri, T., Houde, M., et al.\ 2014, \mnras, 438, 663
\bibitem[Indriolo \& McCall(2012)]{Indriolo12} Indriolo, N., \& McCall, B.~J.\ 2012, \apj, 745, 91
\bibitem[Kulsrud \& Pearce(1969)]{KP69} Kulsrud, R., \& Pearce, W. L. 1969, \apj, 156, 445 
\bibitem[Lam et al.(2019)]{Lam19} Lam, K.~H., Li, Z.-Y., Chen, C.-Y., et al.\ 2019, \mnras, 489, 5326
\bibitem[Lehmann \& Wardle(2016)]{LehmannWardle16} Lehmann, A., \& Wardle, M.\ 2016, \mnras, 455, 2066
\bibitem[Lesur et al.(2014)]{Lesur14} Lesur, G., Kunz, M.~W., \& Fromang, S.\ 2014, \aap, 566, A56
\bibitem[Li et al.(2008)]{LiPS08} Li, P.~S., McKee, C.~F., Klein, R.~I., et al.\ 2008, \apj, 684, 380
\bibitem[Li \& Houde(2008)]{LiHoude08} Li, H.-. bai ., \& Houde, M.\ 2008, \apj, 677, 1151
\bibitem[Li et al.(2014)]{LiHB14} Li, H.-B., Goodman, A., Sridharan, T.~K., et al.\ 2014, Protostars and Planets VI, 101
\bibitem[Lizano \& Shu(1989)]{LizanoShu1989} Lizano, S., \& Shu, F.~H.\ 1989, \apj, 342, 834
\bibitem[Mac Low et al.(1995)]{MacLow95} Mac Low, M.-M., Norman, M.~L., Konigl, A., et al.\ 1995, \apj, 442, 726
\bibitem[Mac Low \& Smith(1997)]{MacLow97} Mac Low, M.-M., \& Smith, M. D. 1997, \apj, 491, 596
\bibitem[Masson et al.(2016)]{Masson16} Masson, J., Chabrier, G., Hennebelle, P., et al.\ 2016, \aap, 587, A32
\bibitem[McKee et al.(2010)]{McKee10} McKee, C. F., Li, P. S., \& Klein, R. I. 2010, \apj, 720, 1612
\bibitem[McKee \& Ostriker(2007)]{MO07} McKee, C.~F., \& Ostriker, E.~C.\ 2007, \araa, 45, 565
\bibitem[Mellon \& Li(2009)]{MellonLi09} Mellon, R.~R., \& Li, Z.-Y.\ 2009, \apj, 698, 922
\bibitem[Mouschovias(1978)]{Mouschovias1978} Mouschovias, T.~C.\ 1978, in IAU Colloq. 52, Protostars and Planets: Studies of Star Formation and of the Origin of the Solar System (Tucson, AZ: Univ. Arizona Press), 209
\bibitem[Mestel \& Spitzer(1956)]{MS1956} Mestel, L., \& Spitzer, L.\ 1956, \mnras, 116, 503
\bibitem[Miura et al.(2012)]{Miura12} Miura, R.~E., Kohno, K., Tosaki, T., et al.\ 2012, \apj, 761, 37
\bibitem[Mouschovias(1979)]{Mouschovias79} Mouschovias, T.~C.\ 1979, \apj, 228, 475
\bibitem[Nakano \& Nakamura(1978)]{Nakano1978} Nakano, T., \& Nakamura, T.\ 1978, \pasj, 30, 671
\bibitem[Nakano et al.(2002)]{Nakano02} Nakano, T., Nishi, R., \& Umebayashi, T.\ 2002, \apj, 573, 199
\bibitem[Nguyen-Luong et al.(2016)]{Nguyen16} Nguyen-Luong, Q., Nguyen, H.~V.~V., Motte, F., et al.\ 2016, \apj, 833, 23
\bibitem[Nishi et al.(1991)]{Nishi91} Nishi, R., Nakano, T., \& Umebayashi, T.\ 1991, \apj, 368, 181
\bibitem[Oishi \& Mac Low(2006)]{Oishi06} Oishi, J.~S., \& Mac Low, M.-M.\ 2006, \apj, 638, 281
\bibitem[Ostriker et al.(1999)]{Ostriker99} Ostriker, E.~C., Gammie, C.~F., \& Stone, J.~M.\ 1999, \apj, 513, 259
\bibitem[Ostriker et al.(2010)]{Ostriker10} Ostriker, E.~C., McKee, C.~F., \& Leroy, A.~K.\ 2010, \apj, 721, 975
\bibitem[Pineau des Forets et al.(1997)]{Pineau97} Pineau des Forets, G., Flower, D.~R., \& Chieze, J.-P.\ 1997, in IAU Symp. 182, Herbig-Haro Flows and the Birth of Stars, ed. B. Reipurth \& C. Bertout (Dordrecht: Kluwer), 199
\bibitem[Riols \& Lesur(2018)]{RiolsLesur18} Riols, A., \& Lesur, G.\ 2018, \aap, 617, A117
\bibitem[Shu et al.(1987)]{Shu1987} Shu, F.~H., Adams, F.~C., \& Lizano, S.\ 1987, \araa, 25, 23\bibitem[Shu(1992)]{Shu} Shu, F. H. 1992, in Physics of Astrophysics, Vol. II, ed. F. H. Shu (Mill Valley, CA: Univ. Science Books),
\bibitem[Smith \& Mac Low(1997)]{SmithML97} Smith, M.~D., \& Mac Low, M.-M.\ 1997, \aap, 326, 801
\bibitem[Spitzer(1956)]{Spitzer1956} Spitzer, L.\ 1956, Physics of Fully Ionized Gases
\bibitem[Stone(1997)]{Stone} Stone, J. 1997, \apj, 487, 271
\bibitem[Stone et al.(2008)]{Stone08} Stone, J.~M., Gardiner, T.~A., Teuben, P., et al.\ 2008, \apjs, 178, 137
\bibitem[Suriano et al.(2018)]{Suriano18} Suriano, S.~S., Li, Z.-Y., Krasnopolsky, R., et al.\ 2018, \mnras, 477, 1239
\bibitem[Suriano et al.(2019)]{Suriano19} Suriano, S.~S., Li, Z.-Y., Krasnopolsky, R., et al.\ 2019, \mnras, 484, 107
\bibitem[Tang et al.(2018)]{Tang18} Tang, K.~S., Li, H.-B., \& Lee, W.-K.\ 2018, \apj, 862, 42
\bibitem[Tielens(2005)]{Tielens05} Tielens, A.~G.~G.~M.\ 2005, The Physics and Chemistry of the Interstellar Medium
\bibitem[Valdivia et al.(2017)]{Valdivia17} Valdivia, V., Godard, B., Hennebelle, P., et al.\ 2017, \aap, 600, A114
\bibitem[van Loo et al.(2009)]{vanLoo09} van Loo, S., Ashmore, I., Caselli, P., et al.\ 2009, \mnras, 395, 319
\bibitem[Vaytet et al.(2018)]{Vaytet18} Vaytet, N., Commer{\c{c}}on, B., Masson, J., et al.\ 2018, \aap, 615, A5
\bibitem[Wardle(1990)]{Wardle} Wardle, M. 1990, \apj, 246, 98
\bibitem[Wardle(1991)]{Wardle1991} Wardle, M.\ 1991, \mnras, 251, 119
\bibitem[Xu \& Li(2016)]{XuLi16} Xu, D., \& Li, D.\ 2016, \apj, 833, 90
\bibitem[Yen et al.(2018)]{Yen18} Yen, H.-W., Zhao, B., Koch, P.~M., et al.\ 2018, \aap, 615, A58
\bibitem[Youdin \& Goodman(2005)]{YG05} Youdin, A., \& Goodman, J. 2005, \apj, 620, 459
\bibitem[Zweibel(1998)]{Zweibel98} Zweibel, E.~G.\ 1998, \apj, 499, 746
\end{thebibliography}
\end{document}